\documentclass[a4paper,12pt]{article}
\usepackage{amsfonts,amsthm,amsmath,amssymb,graphicx}
\usepackage{braket}
\usepackage{bm}
\usepackage[subrefformat=parens]{subcaption}
\usepackage{mathrsfs}
\numberwithin{equation}{section}

\newcommand{\be}{\begin{equation}}
\newcommand{\ee}{\end{equation}}
\newcommand{\Tr}{\text{Tr}}

\newcommand{\ba}{\begin{eqnarray}}
\newcommand{\ea}{\end{eqnarray}}
\newcommand{\mcl}{\mathcal}
\newcommand{\f}{\frac}
\newcommand{\s}{\sqrt}

\newcommand{\ep}{\epsilon}
 
 \def\vep{\varepsilon}

 \def\f {\frac}
 \def\ep{\epsilon}

\begin{document}

\begin{titlepage}
\thispagestyle{empty}

\begin{flushright}
YITP-16-116

\end{flushright}


\begin{center}
\noindent{{\textbf{Scattering effect on entanglement propagation in RCFTs}}}\\
\vspace{2cm}
Tokiro Numasawa $^{1,2}$\vspace{1cm}

{\it
$^1$ Yukawa Institute for Theoretical Physics, \\
Kyoto University, Kyoto 606-8502, Japan\\
}
\vskip 1em
{\it $^2$ Kavli Institute for Theoretical Physics,\\ University of California, Santa Barbara, CA 93117, USA}

\vskip 2em
\end{center}

\begin{abstract}
In this paper we discuss the scattering effect on entanglement propagation in RCFTs. 
In our setup, we consider the time evolution of excited states created by the insertion of many local operators. 
Our results show that because of the finiteness of quantum dimension, entanglement is not changed after the scattering in RCFTs. 
In this mean, entanglement is conserved after the scattering event in RCFTs, which reflects the integrability of the system. Our results are also consistent with the free quasiparticle picture after the global quenches.
\end{abstract}

\end{titlepage}
\newpage
\tableofcontents
\section{Introduction}
AdS/CFT correspondence\cite{Mal}, which is one of the realizations of holographic principle\cite{tHooft,Susskind1}, relates string theory on AdS$_{d+1}$ spacetime to $d$ dimensional conformal field theory (CFT$_d$). 
In the Einstein gravity regime, the number of fields in CFT$_d$ should be very large (large $N$) and the coupling between them should be strong. 
Related to these properties, recently the chaotic nature of holographic CFTs is the focus of attention\cite{SS,RS,MSS,CSSTW,Miyaji}. Out-of-time order correlation function (OTOC), or equally the square of the commutator of operators, is one of the  useful quantities to diagnose the chaotic behavior of many body systems\cite{LO,Kitaev}. 
In chaotic system, we can see the chaotic behavior such as Lyapnov behavior, scrambling and Ruelle resonance\cite{Polchinski,TV}. 
On the other hand, in integrable CFTs such as RCFT, the behavior is different and we cannot see such chaotic behavior\cite{Perlmutter,CNO,GQ}.
The non integrability of boundary theory seems to be important to create black holes in the bulk\footnote{$\mcl{N}$=4 SYM is believed to be integrable at large $N$, but this integrability is broken by the finite $N$ correction or the introduction of thermal background. We thank to P.~Caputa for pointing out this.}\cite{Perlmutter,KMS,MaSt}.

These differences between integrable CFTs and chaotic CFTs can be seen by the time evolution of entanglement in excited states.
For example, let us consider the time evolution of entanglement entropy after the global quench in $1+1 d$ CFTs\cite{CC1}. 
First we consider the entanglement entropy of the single interval.
In this case, the results are universal and depend only on the central charge $c$ of the CFT when time $t$ and the length of interval $L$ is sufficiently large compared to the initial correlation length $\xi$. At early time, entanglement entropy grows linearly and saturates at some time determined by the length $L$. 
This can be explained by the freely propagating quasiparticle model.
On the other hand, the entanglement entropy of disjoint region is not universal\cite{ABGH1}. Let us consider the case of two intervals. 
When the theory is integrable, we find that there is a regime that entanglement entropy decreases. In other words, we can see a dip in the time evolution of entanglement entropy.
This phenomena can be explained by the model of freely propagating quasiparticles. 
On the other hand, in non-integrable theories, such quasiparticle dip becomes smaller. We can think of the size of the quasiparticle dip as a degree of scrambling, which is a quantum information theoretic signature of quantum chaos.
In holographic CFTs, which are the maximally chaotic CFTs \cite{MSS}, the quasiparticle dip vanishes.

We can also see such difference in the time evolution of entanglement entropy in local excited states. 
Consider the excited states that are created by the insertion of local operators on the ground states. 
If the theory is integrable, we can see the propagation of quasiparticles. At the late time, the change of entanglement entropy saturates and the value is given by the entanglement between the propagating quasiparticles\cite{NNT,HNTW,Nozaki,GH}. 
On the other hand, in the case of holographic CFTs, the excess of entanglement entropy does not saturate and grows logarithmically in time\cite{CNT,ABGH2,CSST}.   This growth of entanglement is caused by the chaotic interaction of holographic CFTs and can be seen as a kind of scrambling. 

\begin{figure}
\begin{center}
\includegraphics[width=10cm]{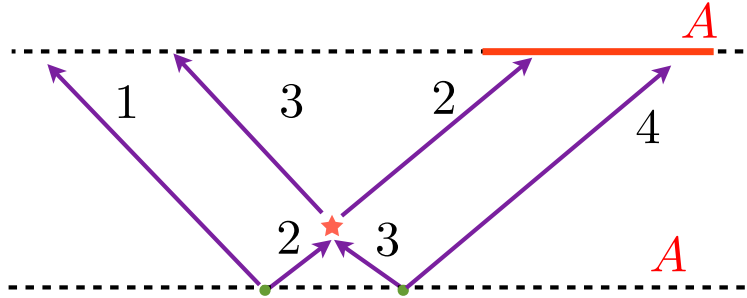}
\caption{The schematic picture of scattering of EPR pairs. The scattering is represented the red star, which corresponds to the unitary matrix $U$ on the Hilbelt space of particle $1$ and $3$.\label{fig:EPRscattering}}
\end{center}
\end{figure}

\begin{figure}
\begin{center}
\includegraphics[width=10cm]{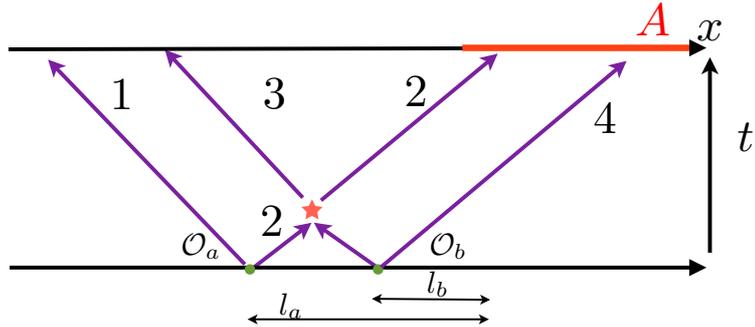}
\caption{The figure of the setup we consider in this paper. 
$\mcl{O}_a$ and $\mcl{O}_b$ are primary operators. The index of operators means the sector of each primary operator. 
At $t=0$, these operators are inserted apart from the entangling surface (in this case actually a point) and entangling quasiparticles are emitted. 
$l_a$ and $l_b$ represent the length from entangling surface.
We consider the case that A is given by the half of space $\{x \in \mathbb{R}|x>0 \}$. \label{fig:QPscattering}}
\end{center}
\end{figure}

This difference of entanglement growth after the excitations depends on the property of interaction (i.e. integrable or chaotic)  of the interaction of the systems. 
Then, how can we see the scattering effect on the propagation of entanglement? This is the motivation of this paper. 
For example, in the paper \cite{CLM} they consider the effect of scattering between two EPR pairs on the propagation of entanglement (Figure \ref{fig:EPRscattering}). 
The initial state is given by the tensor product of two EPR pairs:
\be
\ket{\psi}= \f{1}{1+|\alpha|^2 } (\ket{00}_{12} + \alpha\ket{11}_{12}) \otimes(\ket{00}_{34} + \alpha\ket{11}_{34}),
\ee
where the index of vectors means the label of particles. The initial entanglement entropy between particles $1,3$ and $2,4$ is given by the twice of entanglmenet entropy of EPR pairs $1/\s{1 +|\alpha|^2}(\ket{00}_{12} + \alpha\ket{11}_{12})$.
The scattering effect is given by the action of a unitary matrix $U\in U(4)$ on the Hilbert space of particle 2 and 3. Then the state after the scattering is given by 
\be
\ket{\psi_f} = (\bm{1}\otimes U\otimes \bm{1}) \ket{\psi_i}.
\ee
For general $U$, entanglement entropy between particles $1,3$ and $2,4$ changes after the scattering event. 
In quantum field theory, the scattering effect $U$ should be determined by the Hamiltonian of the system. 
Then, we expect that the scattering effect on the entanglement reflects the property of system, especially the integrability or chaotic nature. 
In this paper, we consider the scattering of local excitations in 2d conformal field theory, especially in RCFT that describes integrable systems. 
In the case of RCFT, we can create the pair of quasi-particles by the action of local operator $\mcl{O}_a$ on the ground state
$\mcl{O}_a\ket{0}$,
where the index $a$ is label of the conformal family that the primary operator belongs to. 
In RCFT, as shown in the paper \cite{HNTW,CO,CGHW}, entanglement between quasi-particles are given by $\log d_a$ where $d_a $ is so called quantum dimension. 
To see the interaction effect on entanglement, first we need to prepare two entangling quasi-particles. This is done by the insertion of two local operators:
\be
\ket{\psi_i}=\mcl{O}_a\mcl{O}_b\ket{0}
\ee 
Then, if we can calculate the entanglement entropy after the scattering, we can see the scattering effect on entanglement (Figure.\ref{fig:QPscattering}). 
This can be done if we follow the time evolution of entanglement entropy of the state $e^{-iHt}\ket{\psi_i} =e^{-iHt} \mcl{O}_a\mcl{O}_b\ket{0}$.
Therefore the problem reduces to the calculation of time evolution of entanglement entropy after the insertion of two local operators. 
We study these problems in this paper.

This paper is organized as follows. 
In section 2 we briefly review the replica method with single local operator\cite{NNT}. 
We also provide an example in RCFT \cite{HNTW} showing how 2d CFT techniques work in our setup. 
In section 3 we show the general formulation with arbitrary number of operators.
In section 4 we apply the formalism in section3 to the calculation of entanglement entropy after the scattering event. 
First we show the examples of two operators insertion in Ising CFT. 
Then we show the general result with arbitrary number of operators in general RCFT.
In section 5 we conclude and comment on the case of chaotic CFTs like CFT with gravity dual.

\section{Review of single operator case}
\subsection{Construction of states}
In this section, we explain the replica method for the excited states by the insertion of local operators.
First we review the insertion of single operator\cite{NNT}\cite{HNTW}\cite{Nozaki}.
The state excited by an local operator is given by
\be
\ket{\Psi(t)} = \s{\mcl{N}}^{-1} e^{- i H t} e ^{- \ep H}\mcl{O}(-l) \ket{0}.\label{op1}
\ee
Here $\mcl{O}(-l)$ is the operator in the Schr\"{o}dinger picture. $\ep$ is a UV cutoff without which the state (\ref{op1}) is not normalizable and 
is not in a Hilbert space. Physically we can interpret this cutoff as a smearing of the excitation to the energy scale given by this.
$\mcl{N}$ is the normalization factor to make the state (\ref{op1}) an unit vector.
The state (\ref{op1}) is understood as the insertion of an local operator 
\be
\mcl{O}(\tau_e,-l) = e^{- \ep H} e^{- i H t} \mcl{O}(-l) e^{i H t }e^{\ep H}. \label{precursor}
\ee
The factor $e^{i H t} e^{\ep H}$ on the right hand side of $\mcl{O}(-l)$ is cancelled after acted on the ground state $\ket{0}$, because $\ket{0}$ is the eigenstate $H\ket{0}=0 $.  The operator (\ref{precursor}) is evolved by complex time $\tau_e = -\ep - it $.
In the calculation, we first use the Euclidean formalism and finally we analytically continue to the complex time. As we will see later, the value of $\ep$ is important to determine the order of insertion of operators \cite{RS}\cite{HJK}.
The same applies to the bra vectors.
The conjugate of (\ref{op1}) becomes as follows:
\be
\bra{\Psi(t)} = \s{\mcl{N}}^{-1} \bra{0} \mcl{O}^{\dagger} (-l) e^{ i H t} e^{- \ep H}. \label{bra}
\ee
We can see this state as the insertion of local operator
\be
\mcl{O}^{\dagger}(\tau_l,-l) = e^{\ep H} e^{-iHt} \mcl{O}^{\dagger}(-l) e^{i H t} e^{- \ep H}.
\ee
where $\tau_l = \ep - it $ is a complex time which is first treated as a Euclidean time and finally analytically continued to complex time. 
In $1+1$ dimension, it is useful to use the holomorphic coordinate. By using the holomorphic coordinate, the density matrix is given by 
\ba
\rho(t) &=& \mcl{N} \cdot e^{-iHt } e^{-\ep H} \mcl{O}(-l) \ket{0} \bra{0} \mcl{O}^{\dagger} (-l) e^{-\ep}e^{iHt} \notag \\
&=&   \mcl{N}\cdot \mcl{O}({\zeta,\bar{\zeta}})\ket{0} \bra{0} \mcl{O}({\zeta',\bar{\zeta}}').
\ea
Here we defined 
\ba
\zeta = -i(\ep+it) -l, \ \ \ \bar{\zeta} = i(\ep+it) -l,  \notag \\
\zeta'= i(\ep - it )-l, \ \ \ \bar{\zeta}'= -i(\ep - it )-l.
\ea
Notes that $\bar{\zeta} \neq -i(\ep - it) -l$, which is the formal complex conjugate of $\zeta$. What we want is the analytical continuation to Lorentzian regime keeping the cut off $\ep$. Therefore the holomorphic coordinates and anti holomorphic coordinates are not related by the complex conjugate after the analytical continuation to complex times.

\subsection{Replica method}
Next we explain the replica method for the local operator excited states.
First we consider how to construct the states or total density matrices by path integral formalism. 
In the ground state cases, we can construct the states (strictly speaking, the wave functionals evaluated at $\phi(\tau=0) = \psi $) by integrating 
from $ - \infty$ to $0$ in Euclidean time\cite{RT}. 
In the local operator excited states, the only difference is to inserting the operator at $ \tau = \tau_e $ where $\tau$ is Euclidean time:
\ba
\braket{\psi|\Psi(t)} &=& \bra{\psi} e^{- i H t} e ^{- \ep H}\mcl{O}(-l) \ket{0} \notag \\
&=&\int_{\tau=-\infty} ^{\phi(\tau =0) = \psi} \mcl{D}\phi \ \mcl{O}(\tau_e ,-l) e ^{- S[\phi]}.
\ea
As we mentioned, we treat the $\tau_e $ as real variable and finally we analytically continue to complex variable, so we first insert the operator at $\tau = \ep$ and then we continue to $\tau _e = -\ep - it$. 
The bra vector also expressed as a integral from $\tau = 0$ to $\tau = \infty$:
\ba
\braket{\Psi(t)|\psi} &=&  \bra{0} \mcl{O}^{\dagger} (-l) e^{ i H t} e^{- \ep H} \ket{\psi} \notag \\ 
&=& \int^{\tau=\infty} _{\phi(\tau =0)=\psi} \mcl{D}\phi \ \mcl{O}(\tau_l ,-l) e ^{- S[\phi]}.
\ea

\begin{figure}
\begin{center}
\includegraphics[width=10cm]{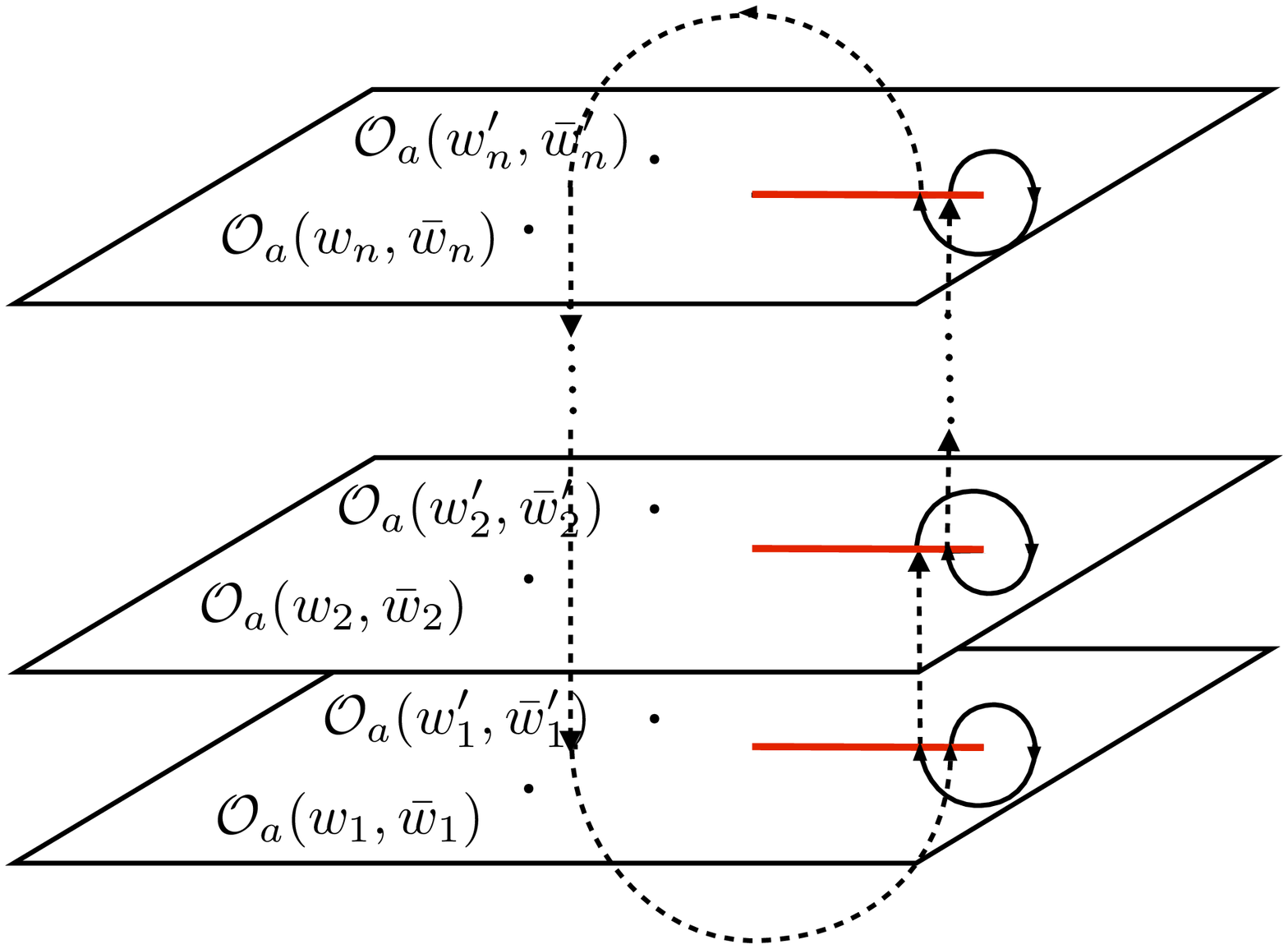}
\caption{ \label{fig:nsheet1op}}
\end{center}
\end{figure}

Next we consider the reduced density matrix. The partial trace of $[\rho]_{\psi \psi'} = \braket{\psi|\Psi}\braket{\Psi|\psi'}$ is realized in path integral formalism \cite{RT} by setting the value of $\psi(x)$ and $\psi'(x)$ for $x \in \bar{A}$ and then integrating by $\psi$ on $\bar{A}$. 
In other words, we make the reduced density matrix by cutting the $A$ on $\tau = 0$ and opposing the boundary condition on each boundary:
\be
[\rho_A]_{\psi_+ \psi_-} = \mcl{N}^{-1} \int _{\tau=-\infty} ^{\tau=\infty} \mcl{D} \phi
\ \mcl{O}(\tau_e,-l) \mcl{O}^{\dagger}(\tau_l,-l) \prod_{x \in A} \delta(\phi(+0,x)- \psi_+(x) ) \delta(\phi(-0,x)-\psi_-(x)) . \label{densitymatrix}
\ee
Here $\mcl{N}$ is the normalization constant. This means that we need to insert the two operators in each sheet, which come from the bra vector and the ket vector.
The normalization constant $\mcl{N}$ is given by tracing out the remaining index of $\rho_A$, which should be $1$ because of the normalization of density matrix. From this , we find that 
the normalization $\mcl{N}$ is given by the unnormalized correlation function or the partition function including the insertion of operators:
\ba
\mcl{N} &=& Z_1 \notag \\ 
&=&\int _{\tau=-\infty} ^{\tau=\infty} \mcl{D}\phi \  \mcl{O}(\tau_e,-l) \mcl{O}^{\dagger}(\tau_l,-l) e ^{- S[\phi]}.
\ea
To find $\Tr \rho_A ^n$, we prepare $n$ copies of (\ref{densitymatrix})
\be
[\rho_A]_{\psi_{1+} \psi_{1-}}[\rho_A]_{\psi_{2+} \psi_{2-}}\cdots[\rho_A]_{\psi _{n+} \psi _{n-}},
\ee
and then take the trace successively. In the path integral language, this procedure corresponds to gluing ${\psi_{i\pm}}$ as $\psi_{i-}(x) = \psi_{(i+1)+}(x) (i=1,2,\cdots, n)$ and 
integrating $\psi_{i+}$. Finally we get the partition function on n-sheeted covering $\Sigma_n$ with $2$ operators on each sheet. 
Then there are $2n$ operators on $\Sigma_n$.  
If we label the coordinates of operator on $i$-th sheet as $x_e^i$ and $x_l^i$, $\Tr\rho_A^n/(\Tr\rho_A)^n$ is given as follows:
\ba
\f{\Tr \rho_A^n}{(\Tr\rho_A)^n} &=&  \f{Z_n}{Z_1^n} \notag \\
&=& \f{\int _{\Sigma_n} \mcl{D} \phi \ \mcl{O}(x_e^1) \mcl{O}^{\dagger}(x_l^1) \cdots \mcl{O}(x_e^n) \mcl{O}^{\dagger}(x_l^n) e ^{- S[\phi]}}
{\Big(\int_{\Sigma_1} \mcl{D}\phi \  \mcl{O}(\tau_e,-l) \mcl{O}^{\dagger}(\tau_l,-l) e ^{- S[\phi]}\Big)^n}
. \label{ptex}
\ea
This seems to be the correlation function on $n$-sheeted manifold, but isn't exactly equal because (\ref{ptex}) is not divided by the normalization factor or the ground state partition function $\int \mcl{D} \phi \ e^{- S[\phi]}$. 
As we will see below, by subtracting the ground state contribution, we can express the trace of of the $n$-th power of the reduced density matrix in terms of correlation functions. The ground state contribution for R\'enyi entropy is given by 
\ba
S_A^{ground} &=& \f{1}{1-n} \log\f{\Tr (\rho^{ground}_A)^n}{(\Tr \rho_A^{ground})^n} \notag \\
&=&\f{1}{1-n} \log\f{\int_{\Sigma_n} \mcl{D} \phi  \ e^{- S[\phi]}}{(\int_{\Sigma_1} \mcl{D} \phi  \ e^{- S[\phi]})^n} .
\ea
Deducing the above contribution , finally we get 
\ba
\Delta S_A^{(n)}(t) &=&\f{1}{1-n} \log\f{\Tr (\rho_A)^n}{(\Tr \rho_A)^n}  -\f{1}{1-n} \log\f{\Tr (\rho^{ground}_A)^n}{(\Tr \rho_A^{ground})^n} \notag \\ 
&=&\f{1}{1-n} \log\f{\int _{\Sigma_n} \mcl{D} \phi \ \mcl{O}(x_e^1) \mcl{O}^{\dagger}(x_l^1) \cdots \mcl{O}(x_e^n) \mcl{O}^{\dagger}(x_l^n) e ^{- S[\phi]}}{\int_{\Sigma_n} \mcl{D} \phi  \ e^{- S[\phi]}} \notag  \\
&& \ \ \ \ \ \ \ \ \ \ \ \  - \f{1}{1-n} \log \f{( \int _{\Sigma_1} \mcl{D}\phi \  \mcl{O}(\tau_e,-l) \mcl{O}^{\dagger}(\tau_l,-l) e ^{- S[\phi]})^n}{ (\int_{\Sigma_1} \mcl{D} \phi  \ e^{- S[\phi]})^n} \notag \\
&=& \f{1}{1-n} \log \f{\braket{\mcl{O}(x_e^1) \mcl{O}^{\dagger}(x_l^1) \cdots \mcl{O}(x_e^n) \mcl{O}^{\dagger}(x_l^n)}_{\Sigma_n}}{(\braket{\mcl{O}(\tau_e,-l) \mcl{O}^{\dagger}(\tau_l,-l)}_{\Sigma_1})^n}.
\ea
From this, for the calculation of entanglement entropy of this excited states, we only need to calculate the $2n$-point function on $n$-sheeted manifolds $\Sigma_n$ (Figure \ref{fig:nsheet1op}).

\subsection{Example: RCFT case}
As an example, let us consider the above excited states in RCFT\cite{HNTW}. R\'enyi entanglement entropies are given by 
\be
\Delta S_A ^{(n)}(t) = \f{1}{1-n} \log \f{\braket{\mcl{O}_a(w_1,\bar{w}_1)\mcl{O}_a^{\dagger}(w_1',\bar{w}_1')\cdots\mcl{O}_a(w_n,\bar{w}_n)\mcl{O}_a^{\dagger}(w_n',\bar{w}_n')  }_{\Sigma_n}}{\braket{\mcl{O}_a(\zeta,\bar{\zeta)}\mcl{O}_a^{\dagger}(\zeta',\bar{\zeta}')}_{\Sigma_1}^n}.
\ee
We consider the R\'enyi entropies of semi infinite interval $A = \{x>0\}$. Using the conformal maps $w = z^n$, we can get
\ba
\f{\braket{\mcl{O}_a(w_1,\bar{w}_1)\mcl{O}_a^{\dagger}(w_1',\bar{w}_1')\cdots\mcl{O}_a(w_n,\bar{w}_n)\mcl{O}_a^{\dagger}(w_n',\bar{w}_n')  }_{\Sigma_n}}{\braket{\mcl{O}_a(\zeta,\bar{\zeta)}\mcl{O}_a^{\dagger}(\zeta',\bar{\zeta}')}_{\Sigma_1}^n} \notag \\
=\mcl{C}_n \cdot \braket{\mcl{O}_a(z_1,\bar{z}_1)\mcl{O}_a^{\dagger}(z_1',\bar{z}_1')\cdots\mcl{O}_a(z_n,\bar{z}_n)\mcl{O}_a^{\dagger}(z_n',\bar{z}_n')  }_{\Sigma_1} \label{ratio}
\ea
where we defined 
\be
\mcl{C}_n = \Bigg( \frac{16 \ep ^4}{n^4 ((l^2 - t^2+\ep^2)^2+4 \ep^2 t^2} \Bigg)^{ n \Delta_a} \cdot \prod _{i=1}^{n} (z_i \bar{z}_i)^{\Delta_a}(z_i' \bar{z}_i')^{\Delta_a},
\ee
and 
\ba
&&z_j = e^{2\pi i \f{j}{n}}(-i \ep + t -l)^{\f{1}{n}}= e^{2\pi i \f{j- 1/2}{n}}(l - t + i\ep )^{\f{1}{n}},\notag \\
&&z_j '= e^{2\pi i \f{j-1}{n}}(i \ep + t -l)^{\f{1}{n}}= e^{2\pi i \f{j- 1/2}{n}}(l - t - i\ep )^{\f{1}{n}}, \notag \\
&&\bar{z}_j = e^{-2\pi i \f{j}{n}}(i \ep - t -l)^{\f{1}{n}}= e^{-2\pi i \f{j- 1/2}{n}}(l + t - i\ep )^{\f{1}{n}},\notag \\
&&\bar{z}_j '= e^{-2\pi i \f{j-1}{n}}(-i \ep - t -l)^{\f{1}{n}}= e^{-2\pi i \f{j- 1/2}{n}}(l + t + i\ep )^{\f{1}{n}}. 
\ea
$\Delta _a$ that appears in the definition of $\mcl{C}_n$ is the (chiral and anti-chiral) conformal dimension of the primary operator $\mcl{O}_a$. The branch cut of $w^{\f{1}{n}}$ is located on $\text{Re} w < 0 $.

\begin{figure}
\begin{minipage}{0.5\hsize}
\includegraphics[width=7cm]{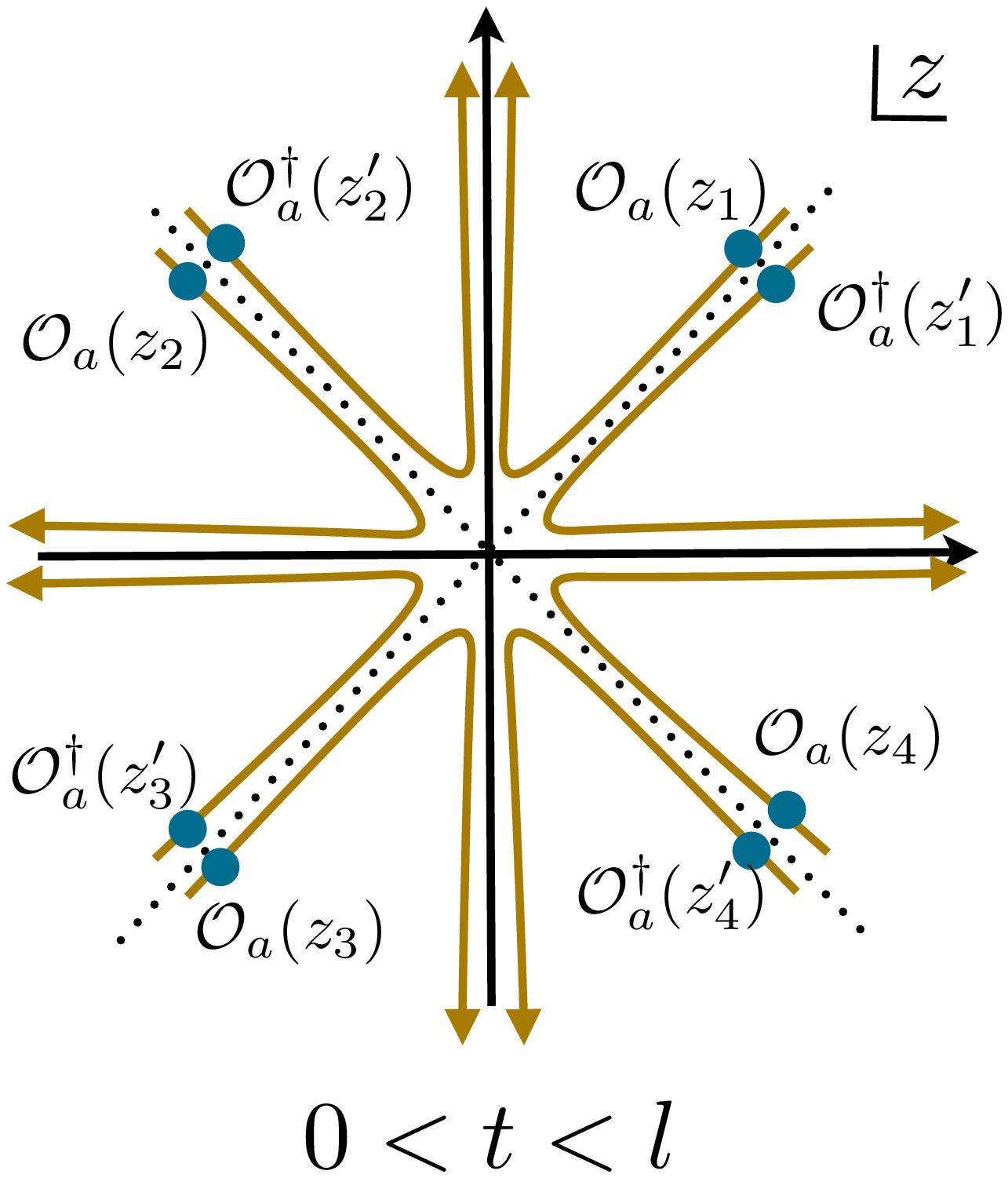}
\end{minipage}
\begin{minipage}{0.5\hsize}
\includegraphics[width=7cm]{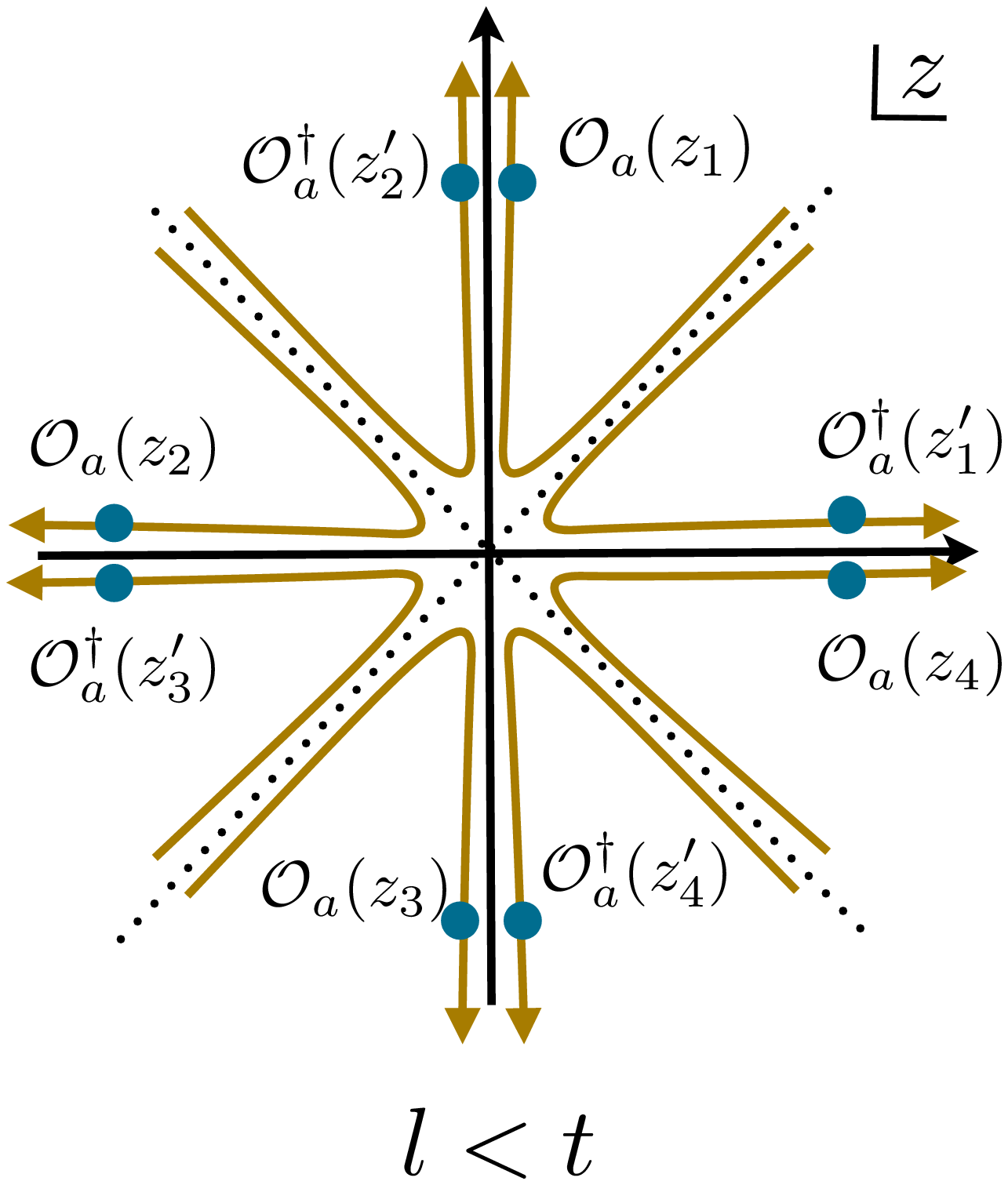}
\end{minipage}
\caption{Configuration of the holomorphic part coordinates of operators on $z$ plane in the $n=4$ case. The blue dots represent the holomorphic coordinate of the local operators. Brown lines are the orbit of coordinates along the time evolution. \label{fig:41}}
\end{figure}

Let us first explain the case of 2nd R\'enyi entanglement entropy. 
In this case, we can express the excess of the R\'enyi entropy using the 4 point function.
Each point is given by 
\ba
&&z_1 = -\s{-i\ep + t -l } = i \s{l - t + i\ep} ,\ \ \ \ \  \ \ \ \bar{z}_1 = -\s{i\ep - t -l } = -i \s{l + t - i\ep}, \notag \\
&&z'_1 = \s{i\ep+ t -l } = i \s{l - t -i\ep} ,\ \ \ \ \ \ \ \ \ \ \ \ \ \bar{z}'_1 = \s{-i\ep - t -l } = -i \s{l + t +i\ep}, \notag \\
&&z_2 = - z_1,  \ \ \ \ \  \ \ \ \bar{z}_1 = -\bar{z}_2 , \notag \\
&&z'_2 = -z'_1, \ \ \ \ \  \ \ \ \bar{z}'_2 = - \bar{z}'_2 .
\ea
It is useful to define the cross ratios $(z,\bar{z})$
\ba
z &=& \f{(z_1 - z'_1)(z_2 - z'_2 )}{(z_1 - z _2 )(z'_1 - z'_2)} = \f{- (l-t )+ \s{(l-t)^2 + \ep ^2 }}{\s{(l-t)^2 + \ep ^2}}, \notag \\
\bar{z} &=& \f{(\bar{z}_1 -\bar{z}'_1)(\bar{z}_2 - \bar{z}'_2 )}{(\bar{z}_1 - \bar{z} _2 )(\bar{z}'_1 - \bar{z}'_2)} = \f{- (l+t )+ \s{(l+t)^2 + \ep ^2 }}{\s{(l+t)^2 + \ep ^2}}. \label{crossratio}
\ea
By using the global conformal transformation, we can fix the 3 points. Thereforfe 4 point function can be expressed as the function of cross ratios $(z,\bar{z})$ as follows:
\be
\braket{\mcl{O}_a(z_1,\bar{z}_1)\mcl{O}_a^{\dagger}(z'_1,\bar{z}'_1)\mcl{O}_a(z_2,\bar{z}_2)\mcl{O}_a^{\dagger}(z'_2,\bar{z}'_2)}_{\Sigma_1} = |(z_1 - z_2)(z'_1 - z'_2)|^{- 4 \Delta _a} G_a (z,\bar{z}).
\ee
Also we can show that the ratio (\ref{ratio}) is expressed as 
\be
\f{
\braket{\mcl{O}_a(w_1,\bar{w}_1)\mcl{O}_a^{\dagger}(w'_1,\bar{w}'_1)\mcl{O}_a(w_2,\bar{w}_2)\mcl{O}_a^{\dagger}(w'_2,\bar{w}'_2)}_{\Sigma_2}
}
{\braket{\mcl{O}_a(\zeta,\bar{\zeta)}\mcl{O}_a^{\dagger}(\zeta',\bar{\zeta}')}_{\Sigma_1}^2} = |z|^{4 \Delta_a }|1-z|^{4 \Delta_a} G_a (z,\bar{z}). \label{ratiowithcr}
\ee
In conformal field theory, we can expand the function $G_a(z,\bar{z})$ using the conformal blocks:
\be
G_a(z,\bar{z}) = \sum_b(C_{aa}^b)^2  f_a(b|z) \bar{f}_a(b|\bar{z}),
\ee
where $b$ runs over all primary fields. In our normalization, conformal blocks behave in $\lim z \to 0$ like
\be
f_a(b|z) = z^{\Delta_b - 2 \Delta_a }(1 + O(z)),
\ee
where $\Delta_b$ is the chiral conformal dimension of the primary operator $\mcl{O}_b$.
From (\ref{crossratio}), at early time $0< t <l$ we find that the cross ratio becomes 
\be
z \simeq \f{\ep^2 }{4 (l-t)^2 }, \ \ \ \ \bar{z} \simeq \f{\ep^2 }{4 (l+t)^2},
\ee
where we used $\simeq$ to denote agreement up to the leading order in $\ep\to 0$ limit. Therefore we obtain $(z,\bar{z}) \to (0,0)$.
This means $G_a(z,\bar{z}) \simeq |z|^{-4 \Delta_a}$, as the dominant contribution arises when $b=0$, which is the identity block.  
This leads to $\Delta S_A^{(2)} = 0$. 
This is expected behavior by the causality that prohibits the propagation faster than the velocity of light.
On the other hand, in the late time $t>l$ we get
\be
z \simeq 1 - \f{\ep^2 }{4 (l-t)^2 }, \ \ \ \ \bar{z} \simeq \f{\ep^2 }{4 (l+t)^2}.
\ee
Therefore we obtain $(z,\bar{z})\to (1,0)$. In this regime, we need to take the fusion transformation, which is the change of basis:
\be
f_a(b|1-z) = \sum_c F_{bc}[a]\cdot f_a(c|z),
\ee
where $F_{bc}[a]$ is a Fusion matrix\footnote{Usually we use the notation $F_{ij}\begin{bmatrix} p &q \\r & s \end{bmatrix}$ for fusion matrix\cite{MS}. Here we denote $F_{ij}\begin{bmatrix} a & a^{\vee} \\a & a \end{bmatrix}$ by $F_{ij}[a]$ for simplicity. }. 
Using this Fusion transformation, in the limit $(z,\bar{z}) \to (1,0)$ we obtain
\be
G_a(z,\bar{z}) \simeq F_{00}[a] \cdot (1-z)^{-2\Delta_a } \bar{z}^{-2 \Delta_a} \label{LCsingular} .
\ee
Therefore we find from (\ref{ratiowithcr}) that 
\be
\Delta S_A ^{(2)}(t) = - \log F_{00}[a] \ \ \ (t>l).
\ee
This is the value of 2nd R\'enyi entropy at late time $t>l$.
 
Next we consider the general $n-$th R\'enyi entropies. 
At early time $t < l$, the configuration of holomorphic part of operators on $\Sigma_1$ is given by the left hand side of Figure\ref{fig:41}. We obtain in $\ep \to 0$ limit
\ba
&&z_j' - z_j \simeq - \f{2i\ep}{n(l-t)} z_j = -\f{2i\ep}{n(l-t)} z_j', \notag \\
&&\bar{z}'_j  -\bar{z}_j \simeq \f{2i\ep}{n(l+t)} \bar{z}_j = \f{2i\ep}{n(l+t)} \bar{z}_j'.
\ea
In this region $2n$-point function is factorized as follows
\be
\braket{\mcl{O}_a(z_1,\bar{z}_1)\mcl{O}_a^{\dagger}(z_1',\bar{z}_1')\cdots\mcl{O}_a(z_n,\bar{z}_n)\mcl{O}_a^{\dagger}(z_n',\bar{z}_n')  }_{\Sigma_1} \simeq \prod_{j = 1} ^{n} \braket{\mcl{O}_a(z_j,\bar{z}_j)\mcl{O}_a^{\dagger}(z_j',\bar{z}_j')}_{\Sigma_1} .
\ee
Therefore we can confirm that the ratio (\ref{ratio}) becomes unity and $\Delta S_A ^{(n)} = 0$.

\begin{figure}
\begin{center}
\includegraphics[width=15cm]{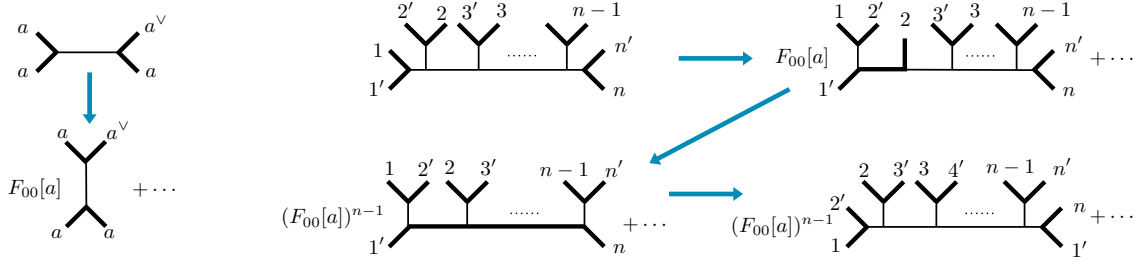}
\end{center}
\caption{The left picture represents the fusion transformation of the 4-point conformal block . The bold connecting two operators shows that we take the OPE of these operators. The thin lines shows the intermediate Identity sector.
The right picture shows the  sequence of fusion transformations to get the R\'enyi entropy $\Delta S_A^{(n)}$, which is a combination of fusion transformations in 4-point formal blocks. \label{fig:fusions}}
\end{figure}

On the other hand, at late time $t>l$, the configuration of holomorphic parts is given by
 the right picture of Figure \ref{fig:41}. 
\ba
&& z_j -z'_{j+1} \simeq -\f{2i\ep}{n(l-t)} z'_{j+1} = -\f{2i\ep}{n(l-t)} z_j, \notag \\
&&\bar{z}_j' -\bar{z}_j \simeq \f{2i\ep}{n(l+t)} \bar{z}_j = \f{2i\ep}{n(l+t)} \bar{z}_j'.
\ea
In order to factorize the $2n$-point functions into $n$ 2-point functions, we need to rearrange the order of the holomorphic coordinates:
\be 
(z_1',z_1)(z_2',z_2) \cdots ( z_{n}',z_{n}) \to (z_1,z_2')(z_2,z_3') \cdots (z_n,z_1'). \label{rearrangement}
\ee
By acting $n-1$ times the fusion transformations as depicted in the Figure \ref{fig:fusions} , we obtain
\be
\braket{\mcl{O}_a(z_1,\bar{z}_1)\mcl{O}_a^{\dagger}(z_1',\bar{z}_1')\cdots\mcl{O}_a(z_n,\bar{z}_n)\mcl{O}_a^{\dagger}(z_n',\bar{z}_n')  }_{\Sigma_1} \simeq (F_{00}[a])^{n-1} \cdot \Bigg[ \prod_{j=1}^{n}(z_j-z_{j+1}')(\bar{z}_j'-\bar{z}_j) \Bigg] .\label{singlefusions}
\ee 
Using this factorization, we find that the ratio (\ref{ratio}) at late time becomes $F_{00}[a]^{n-1} $ . In this way, we obtain the following formula \cite{HNTW}:
\be
\Delta S_A^{(n)}(t)= \log d_a \ \ \ (t>l), 
\ee
where we use the formula $F_{00}[a] = d_a^{-1}$\cite{MS}. 
$d_a$ is so called {\it quantum dimension} and by using modular S matrix $\mcl{S}_{ab}$ this is given by $\mcl{S}_{0a}/\mcl{S}_{00}$. 
This result is interpreted as the entanglement propagation carried by the entangling  quasiparticles that share the entanglement entropy to the amount of $\log d_a$.

Finally we comment on the singularity structure (\ref{LCsingular}) which appears in the calculation of 2nd R\'enyi entropy. 
This type of singularity appears in the setup of entanglement scrambling \cite{ABGH1}.
As mentioned in \cite{ABGH1}, this type of singularity $\bar{z}^{-2 \Delta_a }(1-z)^{-2 \Delta_a}$ does not exist in general CFT. 
Generically, the singularity of $z\to 1$ limit of $f_a(0|z)$ is less singular $(1-z)^{-2 \Delta_{\text{singular}}}$ because of the crossing symmetry $G_a(z,\bar{z}) = G_a(1-z,1-\bar{z})$ \cite{ABGH1}, where $2 \Delta_{\text{singular}}$ is the power of this singularity and satisfy $\Delta_{\text{singular}} \le \Delta_a$.  
In RCFT, because of finiteness of the theory, 
the coefficient $F_{00}[a]$ is always finite and has the power of singularity always given by $2\Delta_a$. 
This finiteness of $F_{00}[a]$  plays an essential role  in the case of many operator insertion case.

\section{Excitations by multiple operators}
In this section, we consider the insertion of many local operators. Naively, the states are given by 
\be
\ket{\Psi(t)} \overset{?}{=} \mcl{O}_{a_1}(\tau_e,-l_1) \cdots \mcl{O}_{a_k}(\tau_e,-l_k)\ket{0},
\ee
using the same $\ep$ as the smearing parameter of local operator.
However in this case, we need to take care of the order of operators. 
In our calculation, first the Euclidean correlation functions and then analytically continue to Lorentzian correlation functions. 
In Euclidean signature, the correlation function given by path integral becomes time ordered correlation function in imaginary time $\braket{0|T_{\tau}(\mcl{O}_{a_1}\cdots\mcl{O}_{a_k})|0}$ in the operator formalism, where $T_{\tau}$ denotes the imaginary time ordering.
Therefore the order of operator is determined by the value of imaginary time. 
We assign the different imaginary times $\{ \ep_p \} $ for each operator. 
Then, the leftmost operator corresponds to the smallest value of $\ep_p$, the second operator corresponds to the second smallest, and so on.
We choose the values of $\{ \ep_p \}$ to be $\ep_1 < \ep_2 \dots < \ep_k$ and then the following state is realized by the path integral with the insertion of local operators:
\be
\ket{\Psi(t)} = \s{\mcl{N}}^{-1}\mcl{O}_{a_1}(\tau_e^1,-l_1)\mcl{O}_{a_2}(\tau_e^2,-l_2) \cdots \mcl{O}_{a_k}(\tau_e^k,-l_k)\ket{0}.
\ee
Here $\tau _e^p = -\ep_p - it  $. In terms of holomorphic coordinates, the position $(\zeta_p$,$\bar{\zeta}_p )$ of each operator is given by
\be
\zeta_p = -i (\ep_p+it ) - l_p, \ \ \ \bar{\zeta}_p= i(\ep_p + it ) -l_p.
\ee

\begin{figure}
\begin{minipage}{0.49\hsize}
\includegraphics[width=70mm]{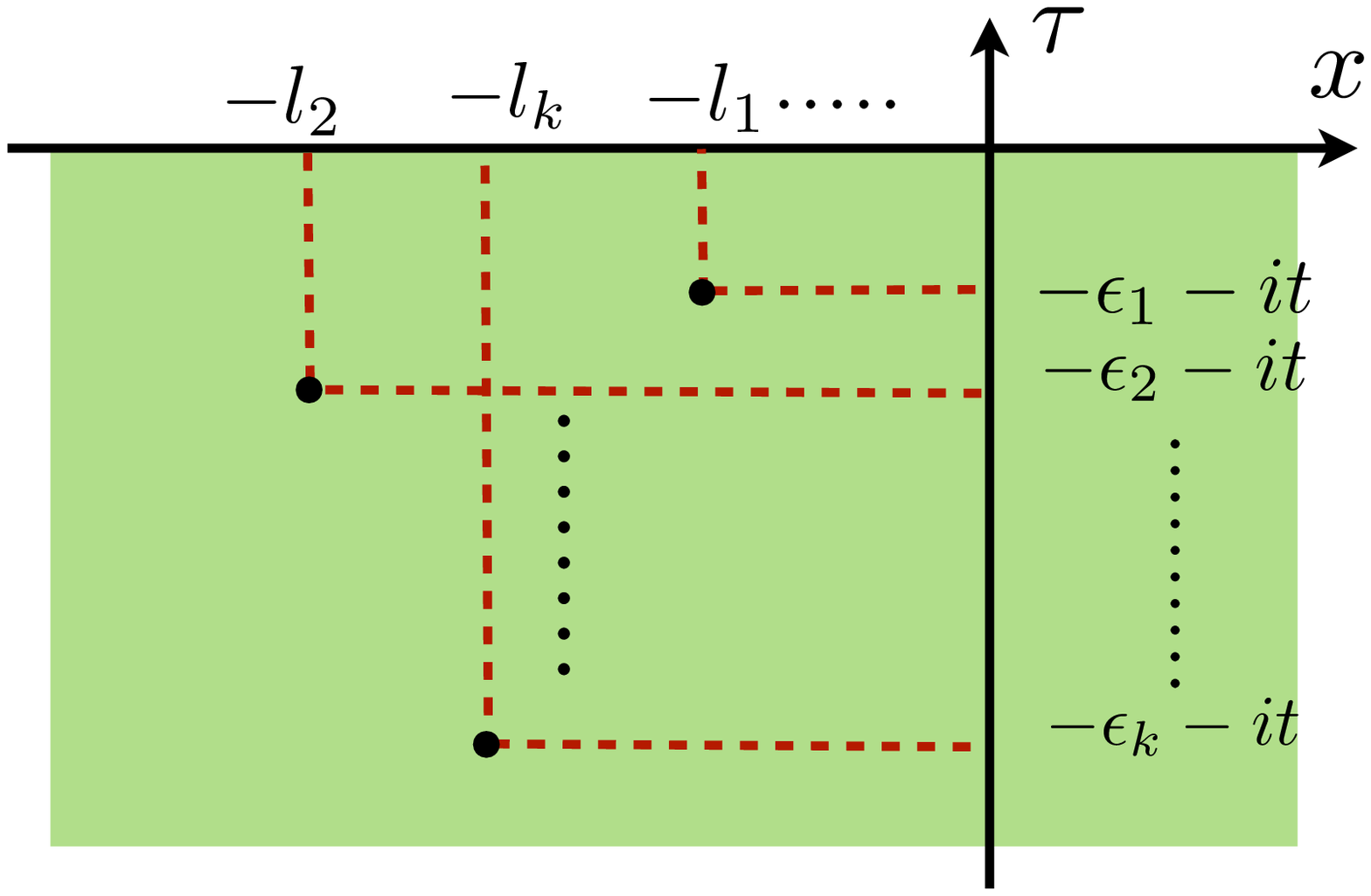}
\end{minipage}
\begin{minipage}{0.49\hsize}
\includegraphics[width=70mm]{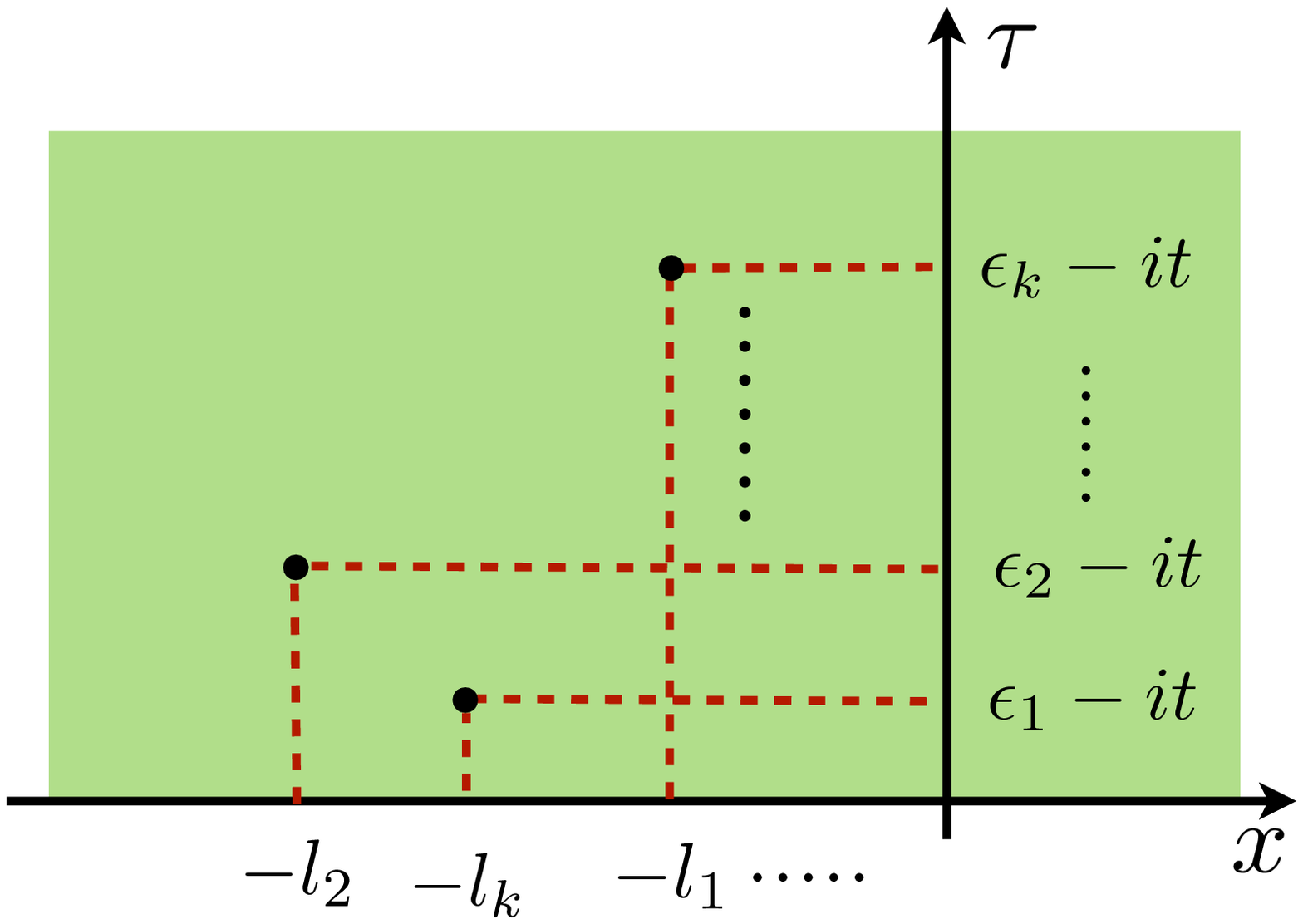}
\end{minipage}
\caption{Path integral representation of the states with $k$ operator insertion.The left picture corresponds to the ket vectors and the right picture corresponds to the bra vectors.  \label{fig:vectormany}}
\begin{center}
\includegraphics[width=10cm]{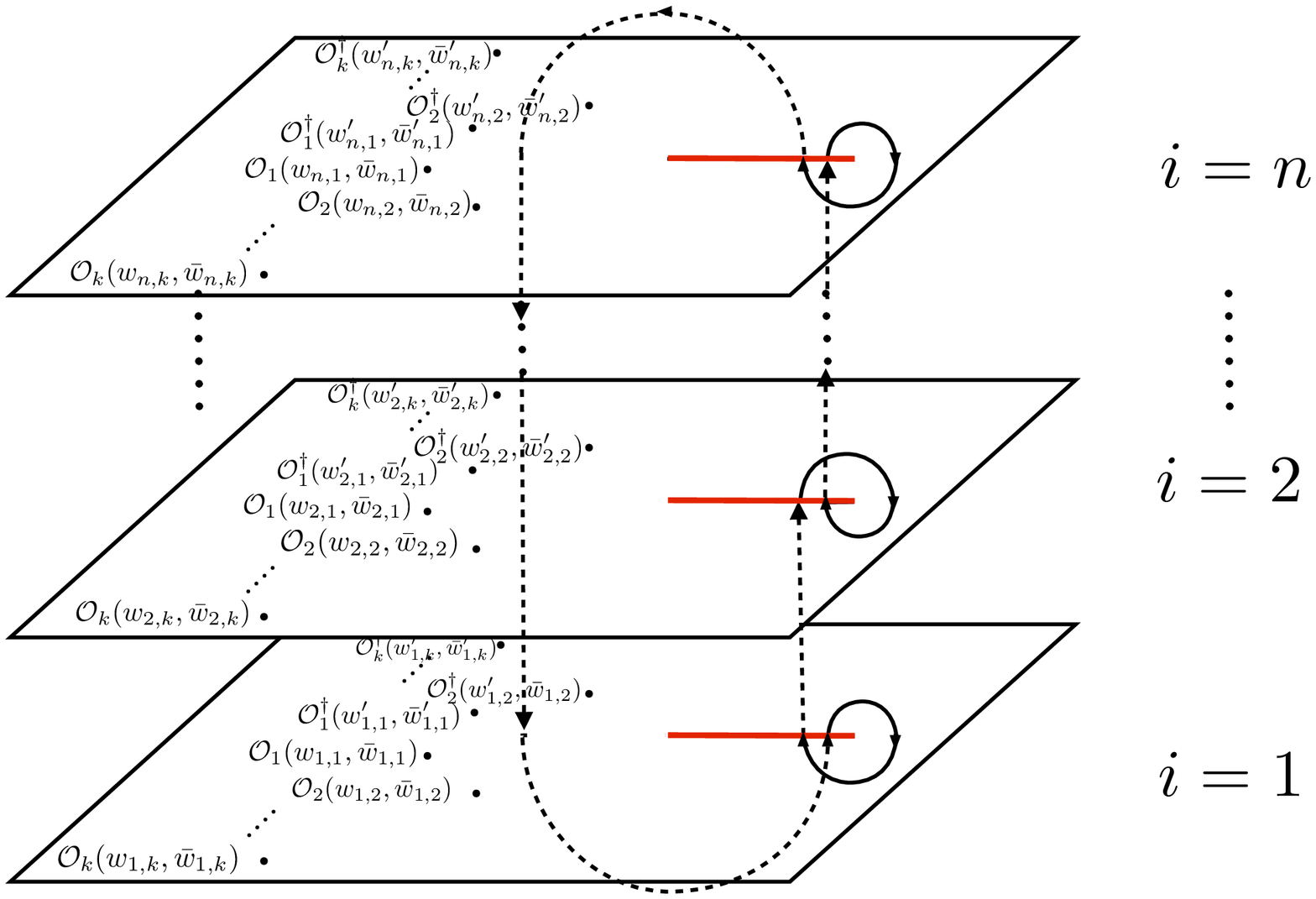}
\caption{$n$-sheeted covering with insertion of $2nk$ operators.\label{fig:coveringmany}}
\end{center}
\end{figure}

\noindent
Similarly, the bra vector is given by 
\ba
\bra{\Psi(t)} &=&\s{\mcl{N}}^{-1} \bra{0}\mcl{O}_k^{\dagger}(\tau_l^k,-l_k)\cdots \mcl{O}_2^{\dagger}(\tau_l^2,-l_2) \mcl{O}_1^{\dagger}(\tau_l^1,-l_1) \notag \\
&=&\bra{0}\mcl{O}_k^{\dagger}(\zeta_k',\bar{\zeta}_k')\cdots \mcl{O}_2^{\dagger}(\zeta_2',\bar{\zeta}_2') \mcl{O}_1^{\dagger}(\zeta_1',\bar{\zeta}_1') ,
\ea
where $\tau_l^p = \ep _p - it$ and each holomorphic coordinate $(\zeta'_p $, $\bar{\zeta}'_p)$ is given by
\be
\zeta'_p = i (\ep_p-it ) - l_p, \ \ \ \bar{\zeta}_p'= -i(\ep_p - it ) -l_p .
\ee
Pictorially, we can represent the bra vector and ket vector as Figure \ref{fig:vectormany}.
In the same manner with the single operator insertion, we can express the R\'enyi entanglement entropy using correlation function:
\be
\Delta S_A^{(n)}(t) = \f{1}{1-n}\log \f{\braket{\prod_{i=1}^{n}\prod_{p=1}^{k}\mcl{O}_{a_p}(w_{i,p},\bar{w}_{i,p})\mcl{O}^{\dagger}_{a_p}(w'_{i,p},\bar{w}'_{i,p})}_{\Sigma_n}}{\braket{\prod_{p=1}^{k}\mcl{O}_{a_p}(\zeta_p,\bar{\zeta}_p)\mcl{O}^{\dagger}_{a_p}(\zeta_p',\bar{\zeta}_p')}_{\Sigma_1}^n}.
\ee

The coordinates $w_{i,p}$ and $w'_{i,p}$ are given by the Figure \ref{fig:coveringmany}\footnote{Here we omitted the explicit expression of coordinate on $\Sigma_n$ because we  only need the coordinate after the conformal map to $\Sigma_1$ and don't need them.}. More explicitly, using the conformal map from $n$-sheeted surface to 1-sheet surface  $f:\Sigma_n \to \Sigma_1$ (here we assume the transformation is given by $z^n = g(w)$ where $z \in \Sigma _1$ and  $ w \in \Sigma_n$ and $g(w)$ is a single-valued holomorphic function), we can express the coordinates $w_{i,p}$ and $w'_{i,p}$ as 
\ba
&&z _{j,p} = f (w_{j,p}) = e^{\f{2 \pi i}{n}j} \sqrt[n]{g(\zeta_p)}, \ \ \ \bar{z} _{j,p} = \bar{f} (\bar{w}_{j,p}) = e^{-\f{2 \pi i}{n}j} \sqrt[n]{\bar{g}(\bar{\zeta}_p)}, \notag \\
&&z _{j,p}' = f (w_{j,p}') = e^{\f{2 \pi i}{n}j} \sqrt[n]{g(\zeta_p')}, \ \ \ \bar{z} _{j,p}' = \bar{f} (\bar{w}_{j,p}') = e^{-\f{2 \pi i}{n}j} \sqrt[n]{\bar{g}(\bar{\zeta}_p')}. \label{renyim}
\ea
In this paper, we only consider the region $A$ is either an infinite half line $A = \{x > 0\}$ or an interval $A = \{0 < x < L \}$. In the former case , the function $g$ is given by $g(w) = w$ . In latter case, conformal case, conformal map $f$ is given by $g(w) = \f{w}{w-L} $.

\section{Many operators excitations in RCFT}
\subsection{Example: 2nd R\'enyi entropy in Ising CFT}
Before considering the general case, first we consider the case of Ising model, which is the first entry of minimal models. This theory has 3 primary operators: Identity $\mathbb{I}$, spin operator $\sigma(z,\bar{z})$ and energy operator $\vep(z,\bar{z})$. The quantum dimension of each operator is given by $d_{\mathbb{I}}=1$, $d_{\sigma }= \s{2}$ and $d_{\vep} = 1$.

\begin{figure}
\begin{center}
\includegraphics[width=17cm]{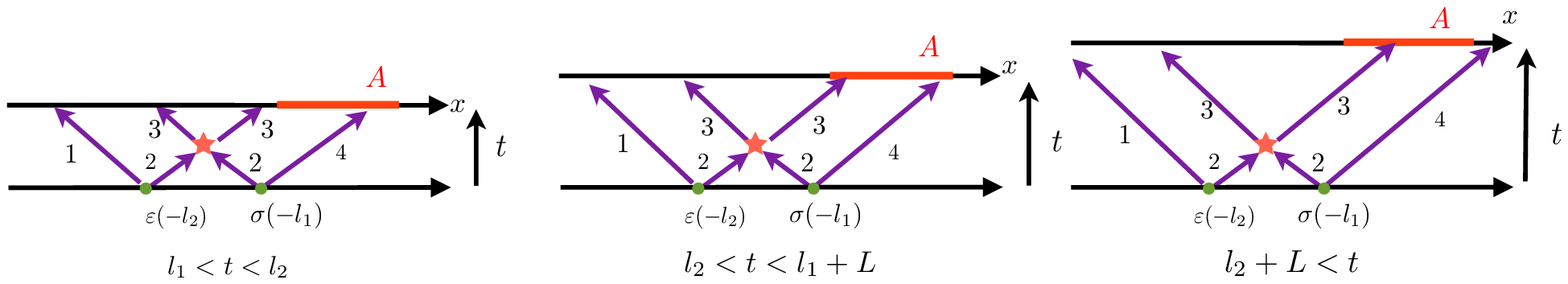}
\caption{\label{fig:es}}
\end{center}
\end{figure}

\begin{figure}
\begin{minipage}{0.49\hsize}
\includegraphics[width=70mm]{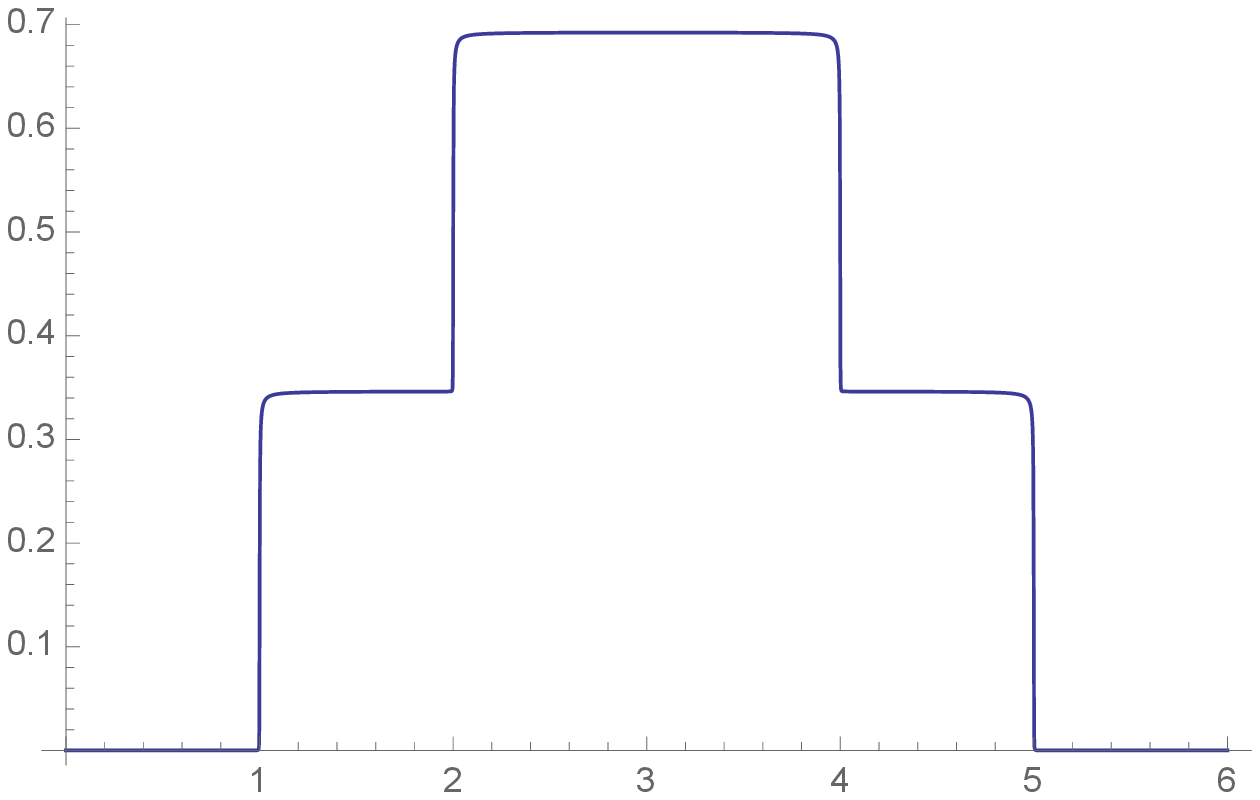}
\end{minipage}
\begin{minipage}{0.49\hsize}
\includegraphics[width=70mm]{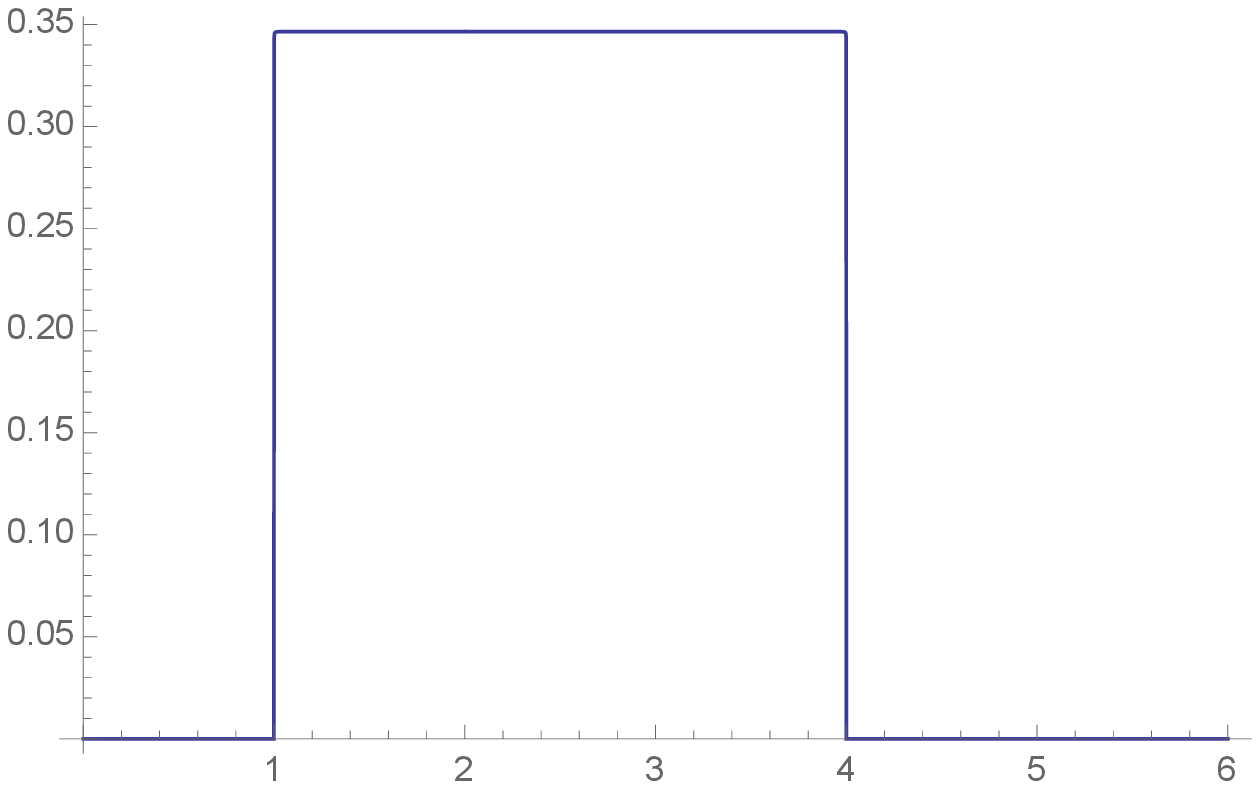}
\end{minipage}
\caption{These are the plots of $2$nd R\'enyi entropy. The vertical line is the difference of R\'enyi entropy from ground state $\Delta S_A^{(2)}$ and the horizontal line is time $t$. The left graph is the time evolution of $\sigma(-l_1)\sigma(-l_2)\ket{0}$ and the right graph is that of $\vep(-l_1)\sigma(-l_2)\ket{0}$. We put $l_1=2$, $l_2=1$ and $L=3$, therefore in the region $2<t<4$ we can see the scattering effect. In the left graph, we put $\ep_1 = 0.001, \ep_2 = 0.002$. In the right graph, we put $\ep_1=0.00001, \ep_2=0.00002$. \label{fig:numerics}}
\end{figure}

In the case of Ising CFT, we can calculate all correlation functions using bosonization technique\cite{Yellowbook}. 
The correlation functions of $2n$ spin operators and $p$ energy operators on a plane are given by
\be
\braket{\sigma(1) \cdots\sigma(2n) \vep (2n+1)\cdots\vep(2n+p)}^2  
= (-1)^p2^n\braket{\prod_{i=1}^{2n} \cos \frac{\varphi}{2}(i) \prod_{k = 2n+1}^{2n+p}(\nabla \varphi/2)^2(k)}, \label{ep:isingcorrelator}
\ee
where the correlation function of two bosons is given by $\braket{\varphi(z,\bar{z})\varphi(w,\bar{w})} = -\ln |z-w|^2$. 
We need to take the square root of the rhs of (\ref{ep:isingcorrelator}) and this square root actually gives a nontrivial correlation between operators.
In our case, to calculate 2nd R\'enyi entropy we need to calculate the following correlation functions:
\begin{flalign}
&\ \ \Delta S_A^{(2)}(t)& \notag
\end{flalign}
\nobreak
\setbox0=\vbox{\parindent=0pt$\displaystyle
= \f{\braket{\sigma(w_{1,1},\bar{w}_{1,1})\sigma(w_{2,1},\bar{w}_{2,1})\sigma(w_{1,2},\bar{w}_{1,2})\sigma(w_{2,2},\bar{w}_{2,2})\sigma(w_{1,1}',\bar{w}_{1,1}')\sigma(w_{2,1}',\bar{w}_{2,1}')\sigma(w_{1,2}',\bar{w}_{1,2}')\sigma(w_{2,2}',\bar{w}_{2,2}')}_{\Sigma_2}}{\braket{\sigma(\zeta_1,\bar{\zeta}_1)\sigma(\zeta_2,\bar{\zeta}_2)\sigma(\zeta_1',\bar{\zeta}_1')\sigma(\zeta_2',\bar{\zeta}_2')}_{\Sigma_1}^2}. $}
\noindent\scalebox{.9}[1]{\box0}
\be
\ee
for $\sigma(-l_1)\sigma(-l_2)\ket{0}$ and 
\begin{flalign}
&\ \ \Delta S_A^{(2)}(t)& \notag
\end{flalign}
\nobreak
\setbox0=\vbox{\parindent=0pt$\displaystyle
= \f{\braket{\vep(w_{1,1},\bar{w}_{1,1})\vep(w_{2,1},\bar{w}_{2,1})\sigma(w_{1,2},\bar{w}_{1,2})\sigma(w_{2,2},\bar{w}_{2,2})\vep(w_{1,1}',\bar{w}_{1,1}')\vep(w_{2,1}',\bar{w}_{2,1}')\sigma(w_{1,2}',\bar{w}_{1,2}')\sigma(w_{2,2}',\bar{w}_{2,2}')}_{\Sigma_2}}{\braket{\vep(\zeta_1,\bar{\zeta}_1)\sigma(\zeta_2,\bar{\zeta}_2)\vep(\zeta_1',\bar{\zeta}_1')\sigma(\zeta_2',\bar{\zeta}_2')}_{\Sigma_1}^2}. $}
\noindent\scalebox{.9}[1]{\box0}
\be
\ee
for $\vep(-l_1)\sigma(-l_2)\ket{0}$.
We consider the case that the subsystem $A$ is given by the interval $[0,  L]$. 
In this case, using the conformal map $z^2 = \f{w}{w-L}$, we can map the $2$-sheet surface  $\Sigma_2$ to the one sheet surface $\Sigma_1 = \mathbb{C}$. 
The correlation function of primary operators are transformed as 
\ba
&&\braket{\sigma (w_{1,1},\bar{w}_{1,1})\cdots\sigma(w_{2,2}',\bar{w}_{2,2}')}_{\Sigma_n} \notag \\
 &=& \prod_{i=1}^{2} \prod_{p=1}^{2} \Big| \f{dw_{i,p}}{dz_{i,p}} \Big| ^{-2\Delta _{\sigma}}\Big| \f{dw'_{i,p}}{dz'_{i,p}} \Big| ^{-2\Delta _{\sigma}} \braket{\sigma (z_{1,1},\bar{z}_{1,1})\cdots\sigma(z_{2,2}',\bar{z}_{2,2}')}_{\Sigma_1} \notag \\
&=& \prod_{i=1}^{2} \prod_{p=1}^{2} \Big| \f{z_{i,p}^2-1}{2 z_{i,p} L}  \Big| ^{2\Delta _{\sigma}}\Big| \f{z'_{i,p}{}^2-1}{2 z'_{i,p} L} \Big| ^{2\Delta _{\sigma}}\braket{\sigma (z_{1,1},\bar{z}_{1,1})\cdots\sigma(z_{2,2}',\bar{z}_{2,2}')}_{\Sigma_1}.
\ea
The same holds for $\vep(-l_1)\sigma(-l_2)\ket{0}$.
Combining the above things, we can follow the time evolution of $2$nd R\'enyi entropy directly. We show the results for $ \epsilon(-l_{1})\sigma (-l_{2})\ket{0}$ and $\sigma(-l_1) \sigma(-l_2)\ket{0}$ in Figure \ref{fig:numerics}.
Especially, we find that in the region $l_2 < t < l_1 + L$, in which region we can see the scattering effect, $2$nd R\'enyi entropy is given by 
\be
\Delta S_A^{(2)}(t) \simeq 0.69
\ee
for $\sigma (-l_1) \sigma(-l_2)\ket{0}$. 
This value is quite close to $2\log d_\sigma $, which is the initial value of $2$nd R\'enyi entropy.
In the case of $\sigma (-l_{\sigma}) \epsilon(-l_{\epsilon})\ket{0}$, we find that 
\be
\Delta S_A^{(2)}(t) \simeq 0.346.
\ee
This is also almost the same value with $\log d_{\sigma} + \log d _{\vep}$, which is the same with the initial $2$nd R\'enyi entropy. These results suggest that after the scattering the entanglement (R\'enyi) entropy is not changed in RCFT. 
This is also suggested from the numerical calculation in corresponding spin system\cite{CR}. 
In the next section we will confirm that this statement is true for generic cases.

\subsection{$n$-th R\'enyi entropies of general RCFT}
Next we show that there are no scattering effect for arbitrary $k$(number of operators) and $n$(replica number) in general RCFT. 
In other words, we show that the (R\'enyi) entanglement entropy is given by the sum of the contributions from each operator:
\be
\Delta S_A ^{(n)}(t) = \sum_{p=1}^{j} \log d_{a_p} \ \ \ (l_j < t < l_{j+1}).
\ee
Here we rearranged ${l_p}$ to be $l_1 < l_2 < \cdots < l_k $ and we put $l_{k+1} = \infty$. We also consider the case that $A$ is an half interval $x>0$ for simplicity. Then the ratio included in (\ref{renyim}) is given by
\ba
\f{\braket{\prod_{i=1}^{n}\prod_{p=1}^{k}\mcl{O}_{a_p}(w_{i,p},\bar{w}_{i,p})\mcl{O}^{\dagger}_{a_p}(w'_{i,p},\bar{w}'_{i,p})}_{\Sigma_n}}{\braket{\prod_{p=1}^{k}\mcl{O}_{a_p}(\zeta_p,\bar{\zeta}_p)\mcl{O}^{\dagger}_{a_p}(\zeta_p',\bar{\zeta}_p')}_{\Sigma_1}^n} \notag \\
= \mcl{C}_{n,k}\cdot \braket{\prod_{i=1}^{n}\prod_{p=1}^{k}\mcl{O}_{a_p}(z_{i,p},\bar{z}_{i,p})\mcl{O}^{\dagger}_{a_p}(z'_{i,p},\bar{z}'_{i,p})}{}_{\Sigma_1}. \label{ratiom}
\ea
where we defined 
\be
\mcl{C}_{n,k} = \prod_{p=1}^{k}\Bigg[\Bigg(\f{16\ep_p^4}{n^4((l_p^2-t^2+\ep_p^2)^2+4\ep_p^2 t^2} \Bigg)^{ n \Delta_{a_p}}\cdot \prod_{i=1}^n(z_{i,a} \bar{z}_{i,a})^{\Delta_a}(z_{i,a}'\bar{z}_{i,a}')^{\Delta_{a_p}}\Bigg] \label{coeff}.
\ee
Generalization to the finite interval is straightforward. 
The essential point is that for arbitrary time $t$ the $2nk $-point function factorizes to the product of $2n $-point functions for each $p$
\be
\braket{\prod_{i=1}^{n}\prod_{p=1}^{k}\mcl{O}_{a_p}(z_{i,p},\bar{z}_{i,p})\mcl{O}^{\dagger}_{a_p}(z'_{i,p},\bar{z}'_{i,p})} {}_{\Sigma_1} \simeq \prod_{p=1}^{k}\braket{\prod_{i=1}^{n}\mcl{O}_{a_p}(z_{i,p},\bar{z}_{i,p})\mcl{O}^{\dagger}_{a_p}(z'_{i,p},\bar{z}'_{i,p})}{}_{\Sigma_1},
\ee
in the $\ep_p \to 0$ limit. This factorization essentially comes from the fact that the conformal block $f_a(0|z)$ has the most divergent singularity $(1-z)^{-2\Delta_a}$ because the coefficient of this term $F_{00}[a]$ is always finite in RCFT.
From this factorization, we can use the same argument of single operator excitations case.
First we explain $n=2$ case how these factorization holds in RCFT even in the lorentzian OPE limit, and then we generalize to $n$-th R\'enyi entropies.

\subsubsection{2nd R\'enyi entropy}
In this case, the difference of R\'enyi entanglement entropy is given using the 8-function of $\mcl{O}_a$'s and $\mcl{O}_b$'s. Here we used the $a,b$ as the label of the 
local operators instead of $a_1$ and $a_2$. We also use $a$ and $b$ as the index $p$ to avoid the complicated notation.
From now, we analyze this 8-point function

\bigskip

\setbox0=\vbox{\parindent=0pt$\displaystyle
\braket{\mcl{O}_a(z_{1,a},\bar{z}_{1,a})\mcl{O}_a(z_{2,a},\bar{z}_{2,a})\mcl{O}_b(z_{1,b},\bar{z}_{1,b})\mcl{O}_b(z_{2,b},\bar{z}_{2,b})\mcl{O}_a^{\dagger}(z'_{1,a},\bar{z}'_{1,a})\mcl{O}_a^{\dagger}(z'_{2,a},\bar{z}'_{2,a})\mcl{O}_b^{\dagger}(z'_{1,b},\bar{z}'_{1,b})\mcl{O}_b^{\dagger}(z'_{2,b},\bar{z}'_{2,b})}_{\Sigma_1} .$}
\noindent\scalebox{.9}[1]{\box0}
\be
\ee

In order to avoid the lengthy notation, we sometimes omit the coordinates of operators and represent each operator by $\mcl{O}_a$ instead of $\mcl{O}_a(z_{i,a},\bar{z}_{i,a})$. In this notation, the 8-point function is expressed as 
\be
\braket{\mcl{O}_a\mcl{O}_a\mcl{O}_b\mcl{O}_b\mcl{O}_a^{\dagger}\mcl{O}_a^{\dagger}\mcl{O}_b^{\dagger}\mcl{O}_b^{\dagger}}_{\Sigma_1}.
\ee

\noindent
(a)In the early time $0<t < l_a$, the configuration of operators given by the left picture   of Figure \ref{fig:operatormovement22} and the 8-point function factorizes to the combination of 2-point functions:

\begin{figure}
\begin{minipage}{0.32\hsize}
\centering
\includegraphics[width=40mm]{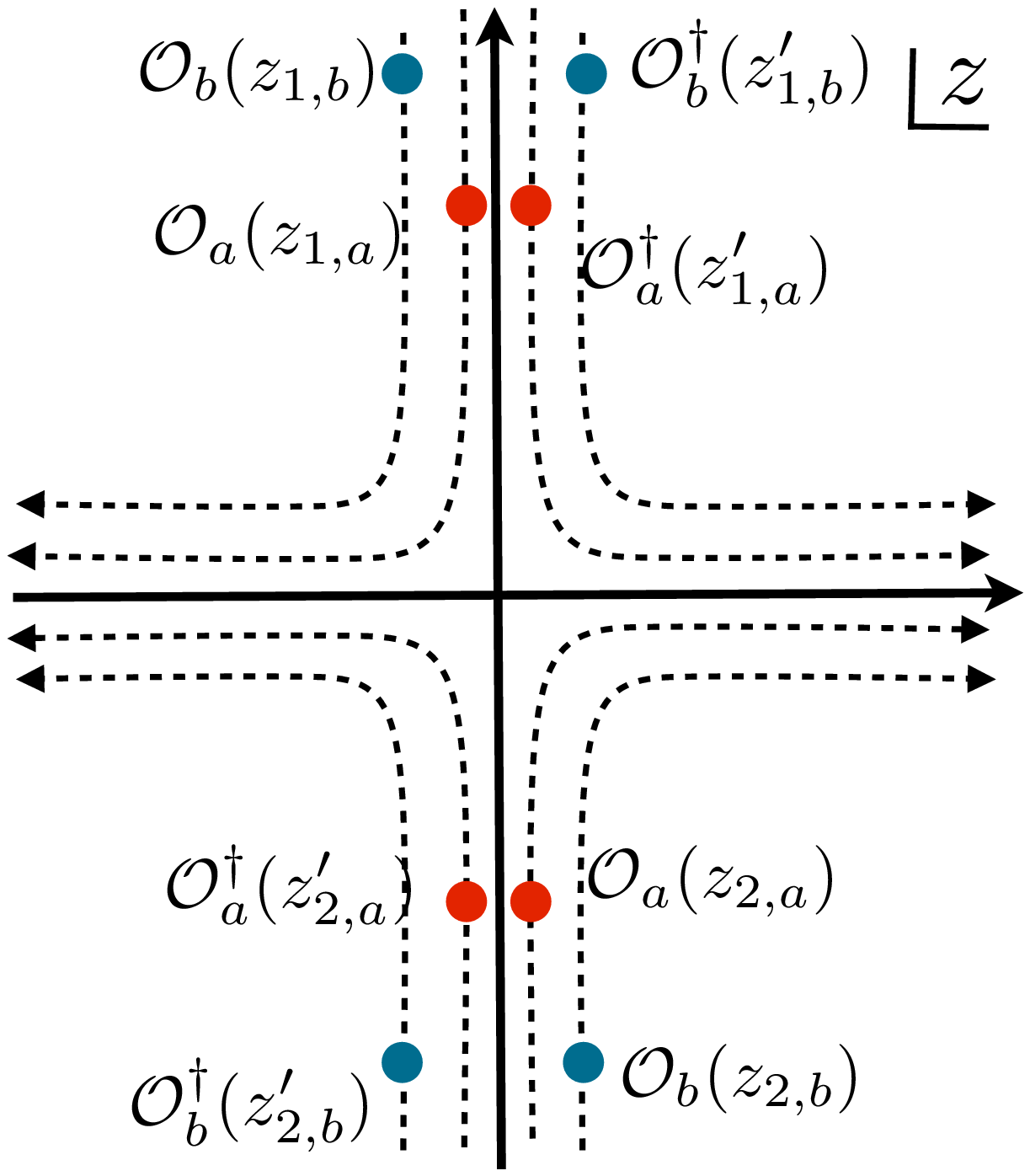}
\subcaption{$0<t<l_a$}
\end{minipage}
\begin{minipage}{0.32\hsize}
\centering
\includegraphics[width=40mm]{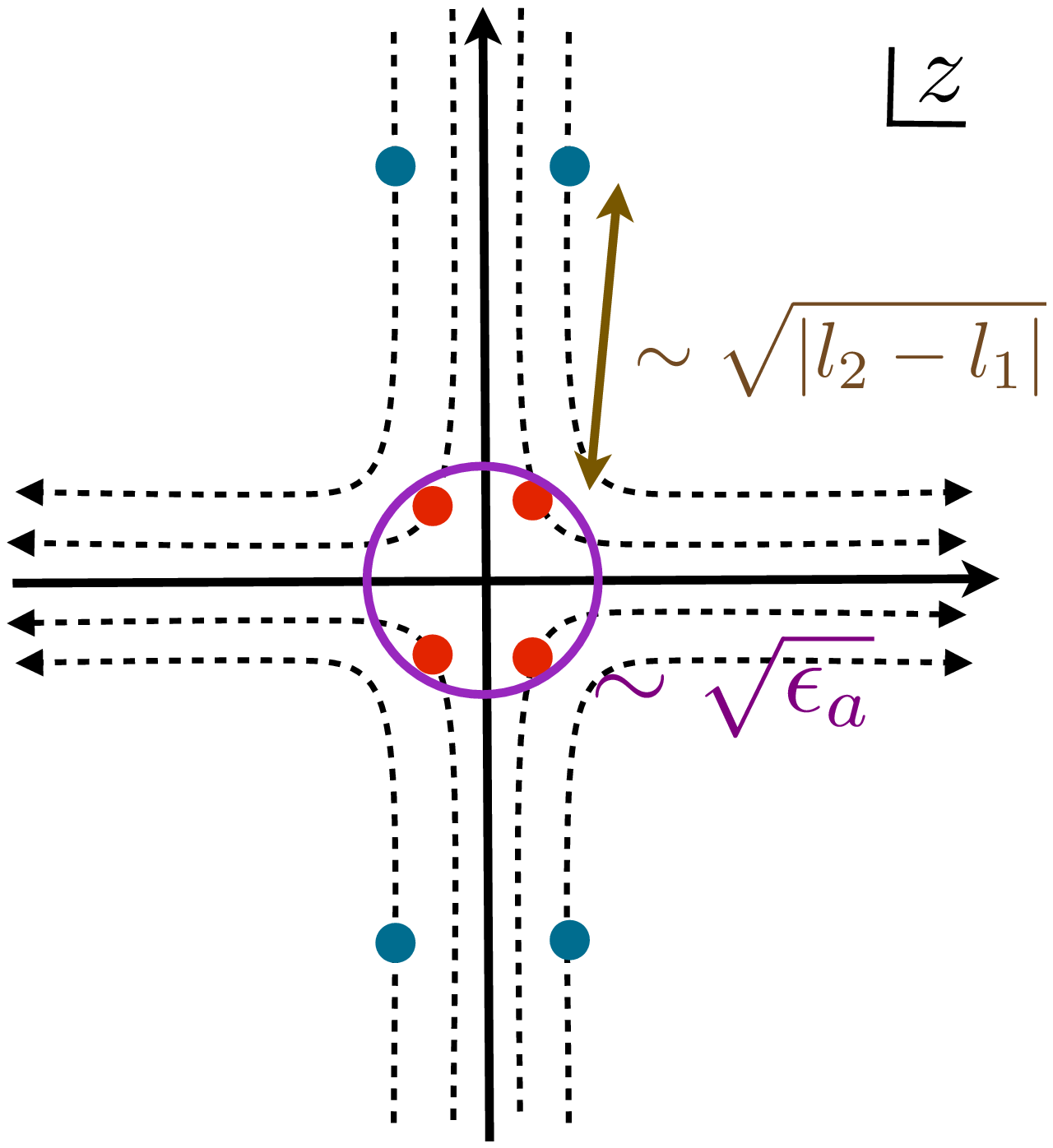}
\subcaption{$t\sim l_a$}
\end{minipage}
\begin{minipage}{0.32\hsize}
\centering
\includegraphics[width=36mm]{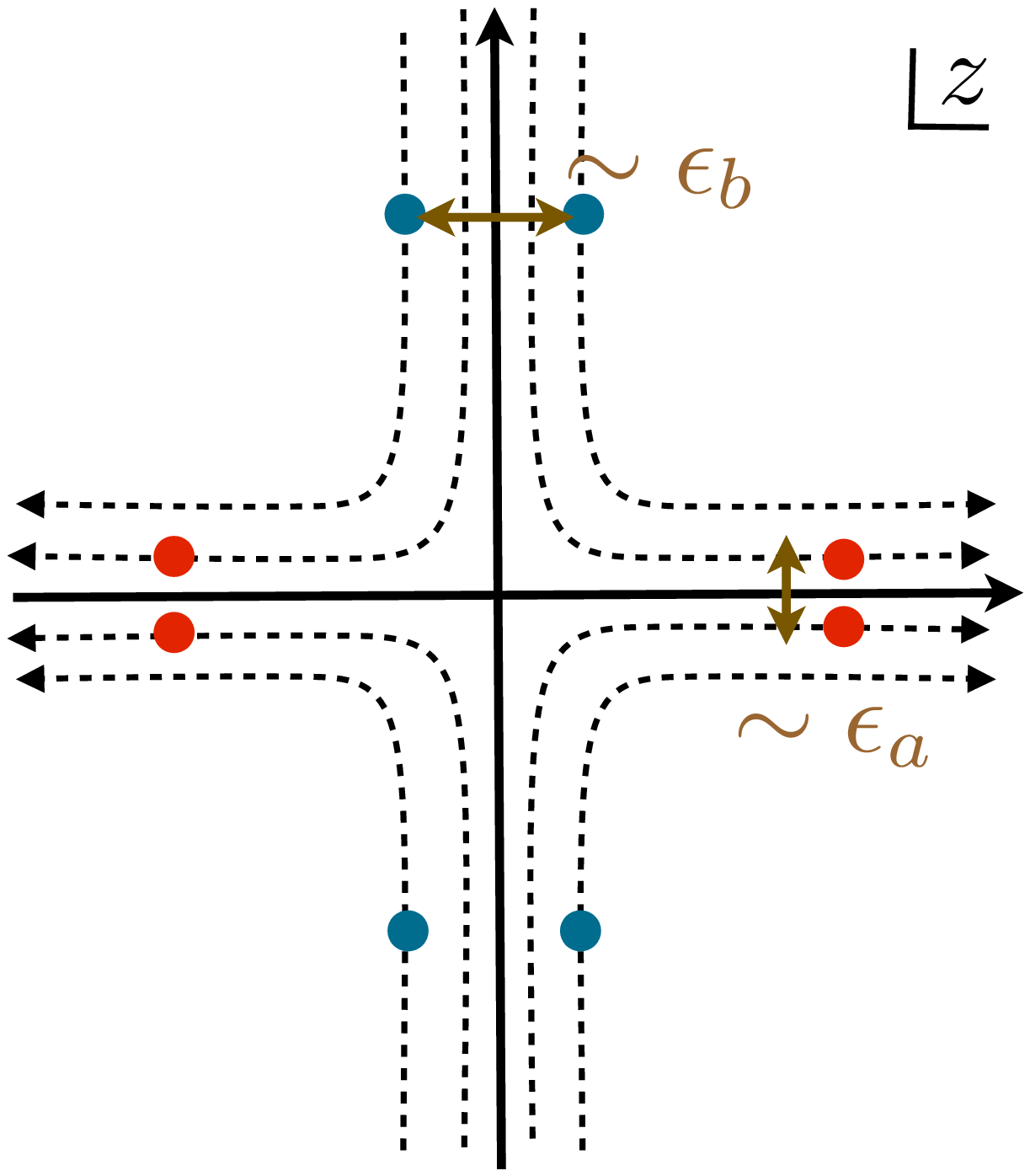}
\subcaption{$l_a<t<l_b$}
\end{minipage}\\
\begin{minipage}{0.32\hsize}
\centering
\includegraphics[width=45mm]{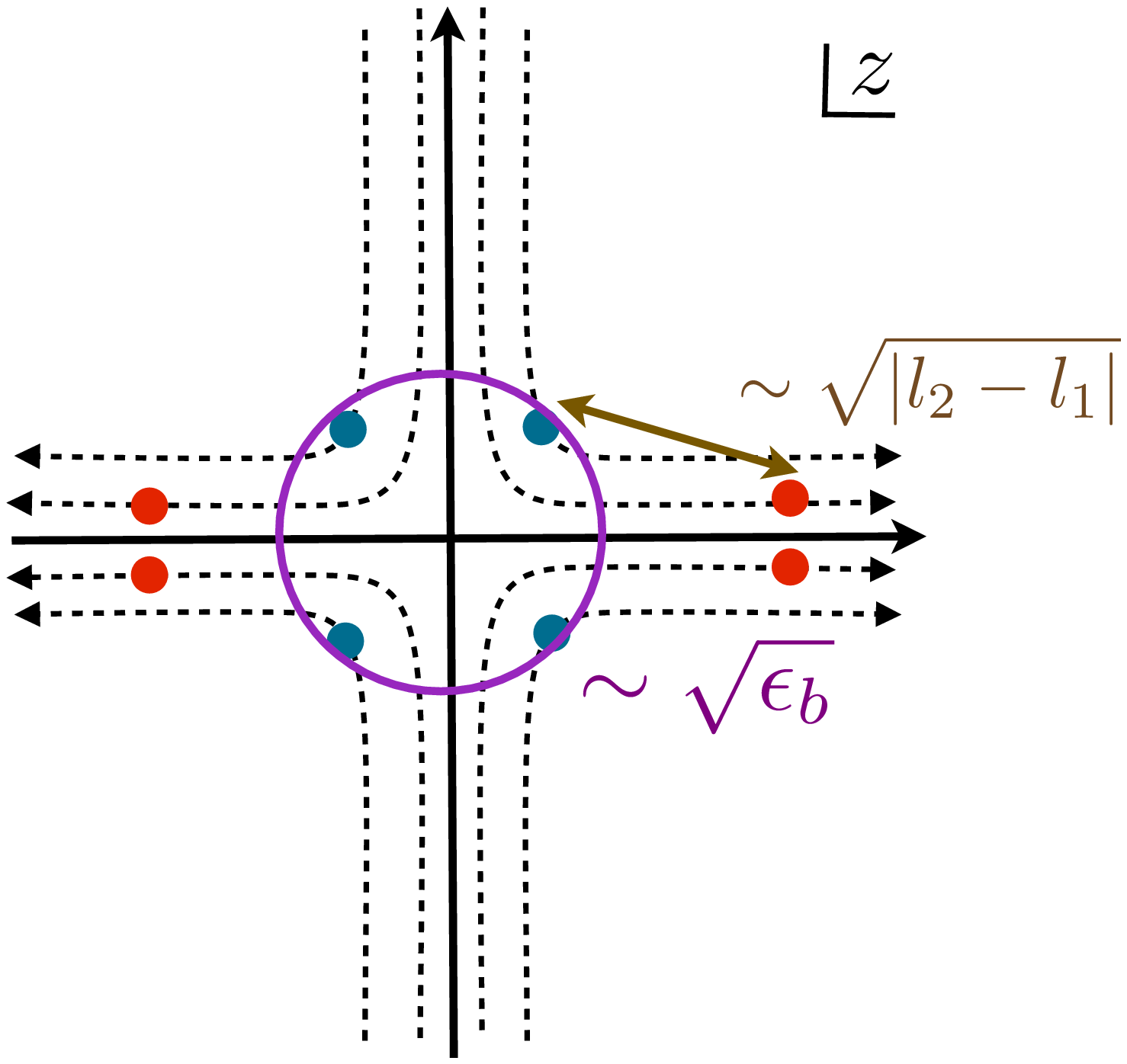}
\subcaption{$t\sim l_b$}
\end{minipage}
\begin{minipage}{0.32\hsize}
\centering
\includegraphics[width=36mm]{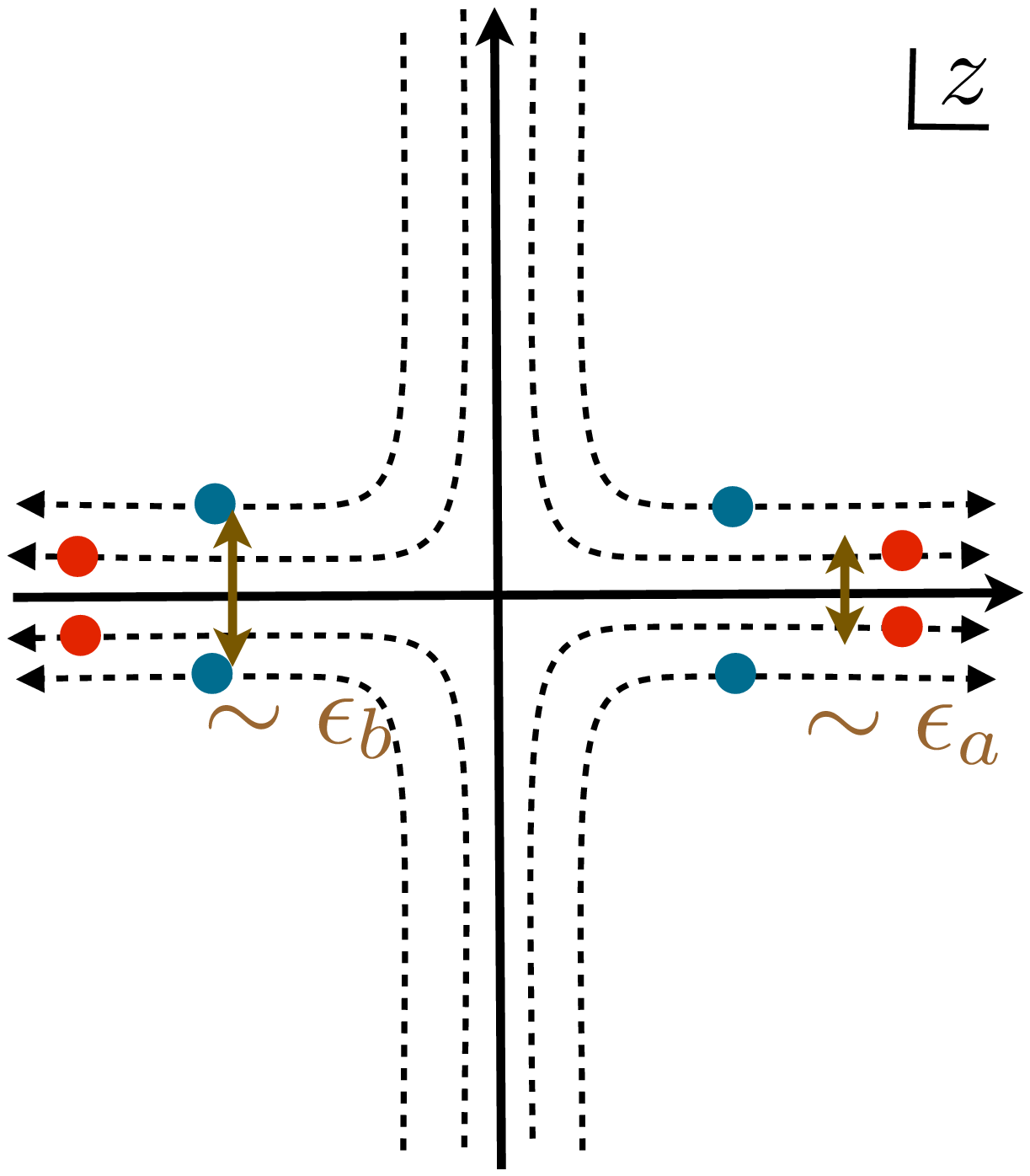}
\subcaption{$l_b<t$}
\end{minipage}

\caption{These pictures show how the chiral parts of local operators move by the time evolution. Local operators labeled by $a$ is represented as the red dots, and the blue dots correspond to the local operators labeled by $b$.  The dotted lines represent the orbits of holomorphic part coordinates of local operators.\label{fig:operatormovement22}} 
\end{figure}

\ba
&&\braket{\mcl{O}_a\mcl{O}_a\mcl{O}_b\mcl{O}_b\mcl{O}_a^{\dagger}\mcl{O}_a^{\dagger}\mcl{O}_b^{\dagger}\mcl{O}_b^{\dagger}}_{\Sigma_1} \notag \\
&&\simeq \braket{\mcl{O}_a(z_{1,a},\bar{z}_{1,a})\mcl{O}_a^{\dagger}(z'_{1,a},\bar{z}'_{1,a})}_{\Sigma_1} \braket{\mcl{O}_a(z_{2,a},\bar{z}_{2,a})\mcl{O}_a^{\dagger}(z'_{2,a},\bar{z}'_{2,a})}_{\Sigma_1} \notag \\
&&\ \ \ \ \ \ \ \ \ \ \ \ \ \  \cdot\braket{\mcl{O}_b(z_{1,b},\bar{z}_{1,b})\mcl{O}_b^{\dagger}(z'_{1,b},\bar{z}'_{1,b})}_{\Sigma_1} \braket{\mcl{O}_b(z_{2,b},\bar{z}_{2,b})\mcl{O}_b^{\dagger}(z'_{2,b},\bar{z}'_{2,b})}_{\Sigma_1} .
\ea

\begin{figure}
\begin{center}
\includegraphics[width=8cm]{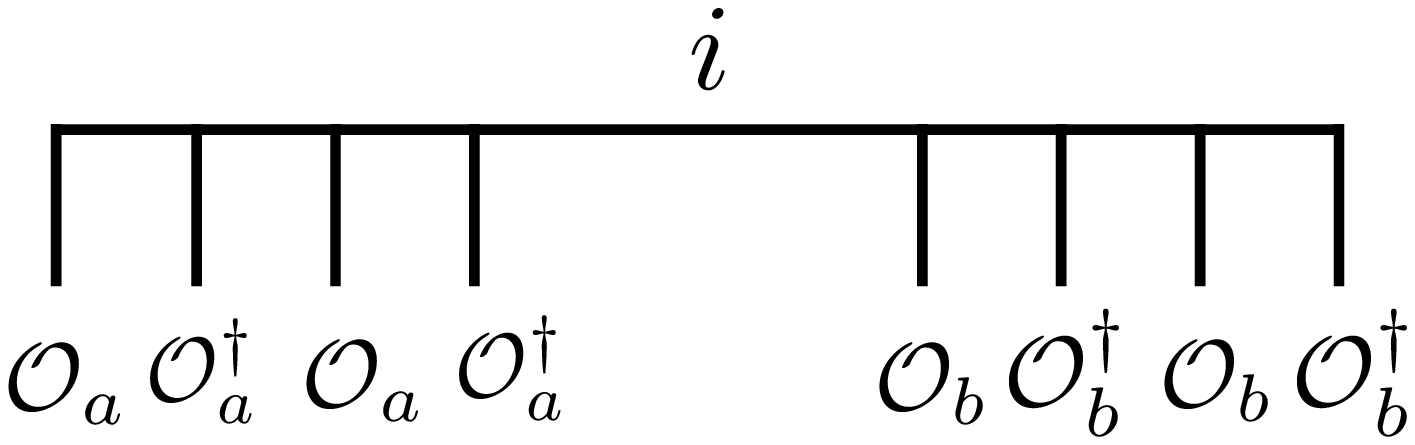}
\caption{\label{fig:intermope}}
\end{center}
\end{figure}

\noindent
(b)Around $t \sim l_a$, the configuration of operators are given by the middle picture of Figure \ref{fig:operatormovement22}. 
In this regime, we can not ignore the existence of $\ep_a$ and the correlation function does not factorize to  the product of 2-point functions. 
However $\ep_a \ll|l_b -l_a| $ still holds. 
Therefore the 8-point function factorizes to the product of 4-point functions:
\ba
&&\braket{\mcl{O}_a\mcl{O}_a\mcl{O}_b\mcl{O}_b\mcl{O}_a^{\dagger}\mcl{O}_a^{\dagger}\mcl{O}_b^{\dagger}\mcl{O}_b^{\dagger}}_{\Sigma_1} \notag \\
&&\simeq \braket{\mcl{O}_a(z_{1,a},\bar{z}_{1,a})\mcl{O}_a^{\dagger}(z'_{1,a},\bar{z}'_{1,a})\mcl{O}_a(z_{2,a},\bar{z}_{2,a})\mcl{O}_a^{\dagger}(z'_{2,a},\bar{z}'_{2,a})}_{\Sigma_1} \notag \\
&&\ \ \ \ \ \ \ \ \ \ \ \ \ \ \cdot\braket{\mcl{O}_b(z_{1,b},\bar{z}_{1,b})\mcl{O}_b^{\dagger}(z'_{1,b},\bar{z}'_{1,b})\mcl{O}_b(z_{2,b},\bar{z}_{2,b})\mcl{O}_b^{\dagger}(z'_{2,b},\bar{z}'_{2,b})}_{\Sigma_1} \notag \\
&&\simeq \braket{\mcl{O}_a(z_{1,a},\bar{z}_{1,a})\mcl{O}_a^{\dagger}(z'_{1,a},\bar{z}'_{1,a})\mcl{O}_a(z_{2,a},\bar{z}_{2,a})\mcl{O}_a^{\dagger}(z'_{2,a},\bar{z}'_{2,a})}_{\Sigma_1}\notag \\ 
&&\ \ \ \ \ \ \ \ \ \ \ \ \ \ \cdot\braket{\mcl{O}_b(z_{1,b},\bar{z}_{1,b})_{\Sigma_1}\mcl{O}_b^{\dagger}(z'_{1,b},\bar{z}'_{1,b})}_{\Sigma_1}\cdot\braket{\mcl{O}_b(z_{2,b},\bar{z}_{2,b})\mcl{O}_b^{\dagger}(z'_{2,b},\bar{z}'_{2,b})}_{\Sigma_1}. \label{eq:8factorization4}
\ea
We explain in detail why this factorization holds.
The correlation function can be expressed using conformal blocks:
\ba
G(z_1,z_2,\cdots ) &=& \sum_{i,j,k,\cdots} a_{ijk\cdots} \mcl{F}_{ijk\cdots}(z_1,z_2,\cdots) \bar{\mcl{F}}_{ijk\cdots}(\bar{z}_1,\bar{z}_2,\cdots) \notag \\
&=& \sum_i \sum_{j,k,\cdots} a_{ijk\cdots} \mcl{F}_{ijk\cdots}(z_1,z_2,\cdots) \bar{\mcl{F}}_{ijk\cdots}(\bar{z}_1,\bar{z}_2,\cdots) \notag \\
&=& \sum_{i} G_i(z_1,\bar{z}_1,z_2,\bar{z}_2,\cdots),
\ea
where $G_i = \sum_{j,k,\cdots}a_{ijk\cdots} \mcl{F}_{ijk\cdots}\bar{\mcl{F}}_{ijk\cdots} $. Therefore we can expand the correlation function with some intermediate states. Here,  we first take the OPE among $\mcl{O}_a$'s and also $\mcl{O}_b$'s separately and then take the remaining OPE(see Figure \ref{fig:intermope}). We take this as $G_{i}$, and we denote this by $\braket{\mcl{O}_a\mcl{O}_a^{\dagger}\mcl{O}_a\mcl{O}_a^{\dagger} \mcl{O}_b\mcl{O}_b^{\dagger}\mcl{O}_b\mcl{O}_b^{\dagger}}_i$.
Because the chiral parts of  $\mcl{O}_a$'s and $\mcl{O}_b$'s are largely separated, we can ignore the the intermediate states. This can be seen, for example, we see the radial direction as time. 
We find the factor $e ^{-(\log{\s{|l_b-l_a|}} - \log{\s{\ep_a}})L_0} = (\ep_a/|l_b-l_a|)^{\f{1}{2}L_0}$, where $L_0$ is the Virasoro operator of level zero. From this, $\braket{\mcl{O}_a\mcl{O}_a^{\dagger}\mcl{O}_a\mcl{O}_a^{\dagger} \mcl{O}_b\mcl{O}_b^{\dagger}\mcl{O}_b\mcl{O}_b^{\dagger}}_i$ is suppressed by the factor $(\ep_a/|l_b-l_a|)^{\f{1}{2}\Delta_i}$.  Therefore we can only take the one that the intermediate state is the identity $\mathbb{I}$ and we can take the fusion transformation among $\mcl{O}_a$'s without taking care of the existence of $\mcl{O}_b$'s .

\begin{figure}
\begin{center}
\includegraphics[width=13cm]{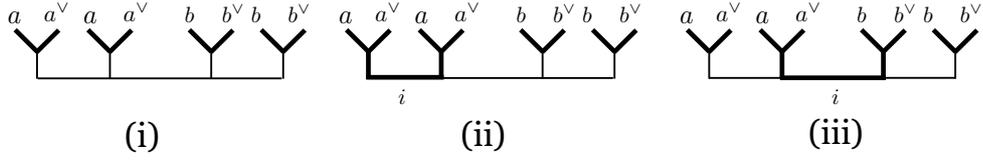}
\caption{(i)leading conformal block (ii)(iii)subleading conformal blocks. (i) gives the contribution of order $\ep_a^{-4\Delta_a}\ep_b^{-4\Delta_b}$ contribution. (ii) gives the order of $\ep_a^{-4\Delta +2\Delta_i} \ep_b^{-4\Delta_b}$ and (iii)  is  the order of $\ep_a^{-4\Delta_a +\Delta_i} \ep_b^{-4\Delta_b+\Delta_i}$. (iii) contains interaction effects through the propagation of the state $\mcl{O}_i$. \label{fig:opeinteraction}}
\end{center}
\end{figure}

\noindent
(c)
In the regime $l_a< t < l_b$, correlation function are factorized but the rearrangement is needed as is the case with single local operator insertion (\ref{rearrangement}). 
In this region, we need rearrangement only among $\mcl{O}_a$'s:
\be
(z_{1,a}',z_{1,a})(z_{2,a}',z_{2,a})(z_{1,b}',z_{1,b})(z_{2,b}',z_{2,b}) \to
(z_{1,a},z_{2,a}')(z_{2,a},z_{1,a}')(z_{1,b}',z_{1,b})(z_{2,b}',z_{2,b}).
\ee
Then, the 8-function is factorized as follows:
\ba
&&\braket{\mcl{O}_a\mcl{O}_a\mcl{O}_b\mcl{O}_b\mcl{O}_a^{\dagger}\mcl{O}_a^{\dagger}\mcl{O}_b^{\dagger}\mcl{O}_b^{\dagger}}_{\Sigma_1} \notag \\
&&\simeq F_{00}[a][(z_{1,a}-z_{2,a}')(z_{2,a}-z_{1,a}')(\bar{z}_{1,a}'-\bar{z}_{1,a})(\bar{z}_{2,a}'-\bar{z}_{2,a})]^{-2\Delta_a}\notag \\
&& \ \ \ \ \ \ \ \cdot \braket{\mcl{O}_b(z_{1,b},\bar{z}_{1,b})\mcl{O}_b^{\dagger}(z'_{1,b},\bar{z}'_{1,b})}_{\Sigma_1}\braket{\mcl{O}_b(z_{2,b},\bar{z}_{2,b})\mcl{O}_b^{\dagger}(z'_{2,b},\bar{z}'_{2,b})}_{\Sigma_1}. \label{leading}
\ea
Because $F_{00}[a]$ is not $0$ but finite in RCFT, the leading term $ 1/(\ep_a^{8 \Delta_a} \ep_b^{8\Delta_b}) $ contained in $\braket{\mcl{O}_a\mcl{O}_a^{\dagger}\mcl{O}_a\mcl{O}_a^{\dagger}}\braket{\mcl{O}_b\mcl{O}_b^{\dagger}\mcl{O}_b\mcl{O}_b^{\dagger}} $ does not vanish under the time evolution of our consideration. The finiteness of $F_{00}[a] = d_a^{-1}$ is important. If the quantum dimension $d_a$ is not finite but infinite, this means the leading term (\ref{leading}) is absent in the 8 point correlator. 
In this case we need to take into account the subleading term contribution, which generically contains the contribution like (ii) or (iii) of Figure \ref{fig:opeinteraction}.
Especially, the conformal block represented by (iii) contains the interaction between $\mcl{O}_a$'s and $\mcl{O}_b$'s through the intermediate state labeled by $i$. 
These nontrivial intermediate states decrese the order of $\ep_a$ and $\ep_b$, and these conformal blocks does not contribute in RCFT.

The contribution from $\mcl{O}_b$'s are cancelled with the factor (\ref{coeff}) and the ratio (\ref{ratiom}) becomes $F_{00}[a]$. 
Therefore 2nd R\'enyi entropy becomes 
\ba
\Delta S_{A}^{(2)}(t) &=& -\log F_{00}[a] \notag \\
&=&\log d_a.   \ \ \ \ \ (l_a < t < l_b)
\ea

\noindent
(d)
Next we consider the time around $t \sim l_b$. 
This is the same with the case $t \sim l_a$ and the configuration of oeprators is given by the right picture of Figure \ref{fig:operatormovement22}. 
In this region we cannot ignore the existence $\ep_b$ and the 8-point function does not  factorizes to the product of 2-point functions. 
However $\ep_b \ll|l_b -l_a|$ still holds and $\mcl{O}_b$'s are decouples from $\mcl{O}_a$'s as is the case with the region $t \sim l_a$. Therefore the 8-point function becomes 
\ba
&&\braket{\mcl{O}_a\mcl{O}_a\mcl{O}_b\mcl{O}_b\mcl{O}_a^{\dagger}\mcl{O}_a^{\dagger}\mcl{O}_b^{\dagger}\mcl{O}_b^{\dagger}}_{\Sigma_1} \notag \\
&&\simeq F_{00}[a][(z_{1,a}-z_{2,a}')(z_{2,a}-z_{1,a}')(\bar{z}_{1,a}'-\bar{z}_{1,a})(\bar{z}_{2,a}'-\bar{z}_{2,a})]^{-2\Delta_a}\notag \\
&&\ \ \ \ \ \ \ \ \ \ \ \cdot \braket{\mcl{O}_b(z_{1,b},\bar{z}_{1,b})\mcl{O}_b^{\dagger}(z'_{1,b},\bar{z}'_{1,b})\mcl{O}_b(z_{2,b},\bar{z}_{2,b})\mcl{O}_b^{\dagger}(z'_{2,b},\bar{z}'_{2,b})}_{\Sigma_1}.
\ea
Therefore we can take the fusion transformation among$\mcl{O}_b$ without taking care of the existence of $\mcl{O}_a$'s.

\noindent
(e)Finally, in the regime $l_b < t$, we need to take into account the rearrangement among $\mcl{O}_b$'s
\be
(z_{1,a},z_{2,a}')(z_{2,a},z_{1,a}')(z_{1,b}',z_{1,b})(z_{2,b}',z_{2,b}) \to 
(z_{1,a},z_{2,a}')(z_{2,a},z_{1,a}')(z_{1,b},z_{2,b}')(z_{2,b},z_{1,b}') .
\ee 
Then,the fusion coefficient $F_{00}[b]$ appears from this rearrangement:
\ba
&&\braket{\mcl{O}_a\mcl{O}_a\mcl{O}_b\mcl{O}_b\mcl{O}_a^{\dagger}\mcl{O}_a^{\dagger}\mcl{O}_b^{\dagger}\mcl{O}_b^{\dagger}}_{\Sigma_1} \notag \\
&&\simeq F_{00}[a] F_{00}[b][(z_{1,a}-z_{2,a}')(z_{2,a}-z_{1,a}')(\bar{z}_{1,a}'-\bar{z}_{1,a})(\bar{z}_{2,a}'-\bar{z}_{2,a})]^{-2\Delta_a}\notag \\
 && \ \ \ \ \ \ \ \ \ \ \ \ \  \ \ \ \ \ \ \ \ \cdot [(z_{1,b}-z_{2,b}')(z_{2,b}-z_{1,b}')(\bar{z}_{1,b}'-\bar{z}_{1,b})(\bar{z}_{2,b}'-\bar{z}_{2,b})]^{-2\Delta_b}. \label{leadingab}
\ea
Note that the leading contribution (\ref{leadingab}) does not vanish because $F_{00}[b]=d_b^{-1}$ is not zero. Otherwise we need to take into account the subleading contributions from $\braket{\mcl{O}_a\mcl{O}_a^{\dagger}\mcl{O}_a\mcl{O}_a^{\dagger} \mcl{O}_b\mcl{O}_b^{\dagger}\mcl{O}_b\mcl{O}_b^{\dagger}}_i$ and the answer will be contain the interaction effect between $\mcl{O}_a$'s and $\mcl{O}_b$'s.
$(z_{1,a}-z_{2,a}')$ 's cancel with $\mcl{C}_{n,k}$ in $\ep_a,\ep_b \to 0$ limit and finally we obtain
\ba
\Delta S_A^{(2)}(t) &=& - \log F_{00}[a] F_{00}[b] \notag \\
&=& \log d_a + \log d_b  \ \ \ \ \   (t < l_b) .
\ea 
Summarizing the above, we find that the time evolution of $2$nd R\'enyi entropy becomes
\be
\Delta S_A ^{(2)}(t) = 
	\begin{cases}
 	0 & 0 < t < l_a \\
 	\log d_a & l_a < t <l_b\\ 
 	\log d_a + \log d_b & l_b < t 
	\end{cases}  .
\ee 

\subsubsection{$n$-th R\'enyi entropies}
The above steps are generalized to general $n$-th R\'enyi entropies $S_A^{(n)}$ with arbitrary number of operators. 
Because in RCFT the fusion coefficients $F_{00}[a_p]=d_{a_p}^{-1}$ are always finite. Therefore , for the same reason with the case of 2nd R\'enyi entropy, in the regime $l_j < t < l_{j+1}$ the $2nk$-point function is factorized as
\ba
&&\braket{\prod_{i=1}^{n}\prod_{p=1}^{k}\mcl{O}_{a_p}(z_{i,p},\bar{z}_{i,p})\mcl{O}^{\dagger}_{a_p}(z'_{i,p},\bar{z}'_{i,p})}{}_{\Sigma_1} \notag \\
&& \simeq \prod_{p=1}^j \Big[ ( F_{00}[a_p] )^{n-1} \Big( \prod_{i=1}^n (z_{i,p}-z_{i+1,p}')(\bar{z}_{i,p}'-\bar{z}_{i,p} ) \Big)^{-2\Delta_{a_p}}\Big] \notag \\
&& \hspace{6cm}  \times  \prod _{p = j+1 }^{k}\prod_{i=1}^{n}  \braket{ \mcl{O}_{a_p}(z_{i,p},\bar{z}_{i,p})\mcl{O}^{\dagger}_{a_p}(z'_{i,p},\bar{z}'_{i,p})}_{\Sigma_1} .
\ea
From this factorization property, in $\ep_p \to 0 $ limit R\'enyi entropy becomes 
\ba
\Delta S_A^{(n)}(t) &=& \f{1}{1-n}\log \Bigg[ \prod _{p=1}^j (F_{00}[a_p] )^{n-1} \Bigg] \notag \\
&=& \sum_{p=1} ^j \log d_{a_p}     \ \ \ \ \ \ \ (l_j<t<l_{j+1}) .
\ea
The time evolution of R\'enyi entropy is given by
\ba
\Delta S_A^{(n)}(t) = 
\begin{cases}
0 & (0 < t < l_1)  \\
\displaystyle\sum _{p = 1}^j \log d_{a_p}  & (l_j < t < l_{j+1}) \\
\displaystyle\sum _{p = 1}^k \log d_{a_p} & (l_k < t) \label{manyrenyin}
\end{cases}.
\ea
This is also conjectured in \cite{CR}.
This is true for any replica number $n$, we can take the analytic continuation $n \to 1 $ and we find that
entanglement entropy is also given by the same value. 
\subsection{scattering effect on entanglement entropy}
From now we will back to the problem of scattering effect on the entangelment propagation. 
Consider the excitation by two operators. 
Before the scattering, entanglement entropy between each quasiparticle pair is given by $\log d_a$. Therefore initial entanglement entropy $S_A^i$ is given by 
\be
S_A^i = \log d_a + \log d_b.
\ee
On the other hand, entanglement entropy after the scattering is given by that of $\Delta S_A $ in the region $t > \max(l_a,l_b)$. 
Therefore, from (\ref{manyrenyin}) we find that entanglement entropy $S_A^f$ after the scattering is given by 
\be
S_A^f = \log d_a + \log d_b .
\ee

Then, we find that the entanglement between particle $1,3$ and $2,4$ does not change after the scattering. 

\begin{figure}
\begin{center}
\includegraphics[width=10cm]{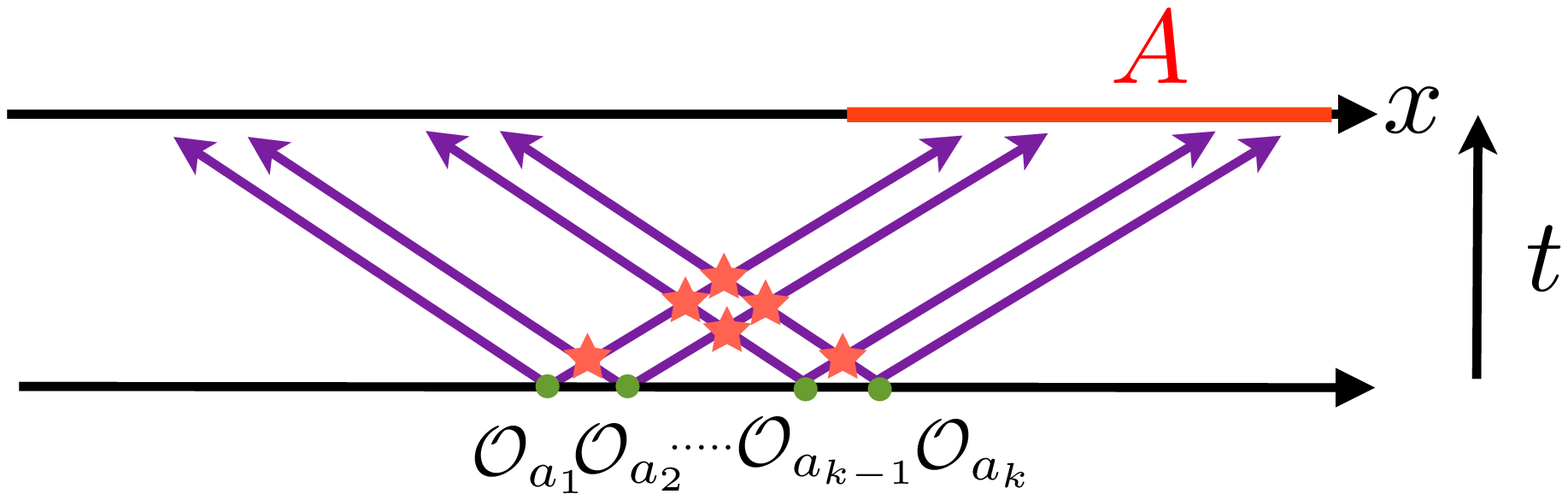}
\caption{\label{fig:mscattering}}
\end{center}
\end{figure}

We can also consider the excitations by many operators like the situation drawn in Figure \ref{fig:mscattering}. 
Also in this case, we the entanglement between left moving quasiparticles and right moving ones quasiparticles are given by
\be
S_A^i = S_A^f = \sum_{p=1}^k \log d_{a_p}.
\ee
Therefore we can conclude that {\it entanglement is conserved } after the scattering in RCFT. This is the main result of this paper.
This is nontrivial statement because RCFTs also include interaction. For example, at first site we can describe minimal models using free boson\cite{Yellowbook} and expect no interaction. 
However in this description there is also a screening operator as an interaction term in the Lagrangian.
Therefore we need to sum various OPE channels and the resulting correlation function is not the same with that of free theory. 
Because the interaction of RCFTs is integrable, which is a special property and leads to the finiteness of $F_{00}[a]=d_a^{-1}$, entanglement entropy does not change after the scattering.

\section{Discussion}

In this paper we have discussed the scattering effect on the entanglement propagation in RCFTs.
In RCFTs, we expect that there should be something special on the effect of entanglement, because the interaction of RCFT is integrable and has the special property like the factorization of many scatterings to two scattering\footnote{Though in CFT we cannot define the asymptotic states and the scattering matrix, we expect that we can define S-matrix by adding a mass term by an integrable deformation.}. 
We showed that the finiteness of quantum dimension $d_a$ leads to the no scattering effect on the entanglement propagation, or the conservation of entanglement under the scattering event.
Our results also support the picture of freely propagating quasiparticles after the global quenches in RCFTs \cite{ABGH1}. 
In RCFTs, after the global quenches quasiparticles are created and freely propagate. 
This leads to the existence of the dip in the time evolution of entanglement entropy of two disjoint intervals. 
Our results of freely propagating quasiparticles are consistent with the existence of the quasiparticle dip\cite{ABGH1} after the global quenches.
It is interesting to note that the finiteness of quantum dimension of twist operator $\sigma_2$ in RCFT\footnote{To be more precise, the twist operator is the primary field in the orbifold theory $\text{CFT}^{\otimes n}/\mathbb{Z}_n$ and quantum dimension is defined in this theory.}  leads to the quasiparticle dip after global quenches\cite{ABGH1}.
It is also interesting that even in the case of many operators insertion, we can extract only the quantum dimension, which is the first column of modular S-matrix. 
On the other hand out-of-time order correlators(OTOC), which is a quantity that can diagnose the quantum chaotic property of many body systems, can extract all the matrix element of modular S-matrix\cite{CNO,GQ}. 
It is interesting problem to construct the setup that can extract all the matrix element of modular S-matrix only using (R\'enyi) entanglement entropy. 
Our results hold in all RCFTs, especially in $W_N$ minimal model, which has and holographic dual with higher spin symmetry\cite{GG}. 
It is interesting to consider the holographic dual of many local operator insertion in these models. Physically, this will be the higher spin analog of black hole collapse\cite{AHRS}.

Finally we comment on the holographic CFTs.
In this paper we concentrated on the RCFT case. Contrary to RCFTs, holographic CFTs ,which describe the gravity, will be quite opposite and believed to be chaotic\cite{SS,RS}. 
The situation is quite different even in the case of single local operator insertion\cite{CNT,ABGH2}.
Actually, the corresponding term to the quantum dimension is divergent at least in the large $c$ limit\cite{CNT}.
Therefore we cannot ignore the subleading terms and  expect that the entanglement propagation is not the naive summation of two excitations. 
These can be expected by considering the holographic dual of a local excitation by insertion of a heavy local operator\cite{ABGH2}. 
The holographic dual of heavy operator insertion is given by a falling particle in AdS and its back reaction\cite{NNTlocal}, which is sometimes called as conformal soliton\cite{FGMP,FHRR}. 
Therefore, the holographic dual of two local operators insertion should be represented by the two falling particles and its back reacted geometry. 
These geometry will not be represented by the naive combination of two falling particles and the metric will take more complicated form.
It is interesting to construct the geometry with two falling particles. 
At least in 2+1 dimension, the geometry is locally AdS$_3$, therefore we may be construct the exact metric as a quotient of AdS$_3$.

\section*{Acknowledgements}

We would like to thank
Masahiro Nozaki,
Kento Watanabe,
Song He,
Wu Zhong Guo,
Shinsei Ryu, Mark Mezei, 
Sunil Mukhi 
and 
Tomonori Ugajin
for valuable conversations and 
expecially Tadashi Takayanagi and Pawel Caputa for reading our draft and giving useful comments.
TN is supported by JSPS fellowships.


\begin{thebibliography}{}
\bibitem{Mal}
J.M.~Maldacena,
``The Large N Limit of Superconformal Field Thheories and Supergravity,''
Adv.Theor.Math.Phys.{\bf2}(1998),
arXiv:hep-th/9711200. 

\bibitem{tHooft}
G.~'t Hooft,
  ``Dimensional reduction in quantum gravity,''
  arXiv:qr-qc/9310026. 
  
\bibitem{Susskind1}
L.~Susskind,
``The World as a hologram,''
J.\ Math.\ Phys.\  {\bf 36}, 6377 (1995)
[arXiv:hep-th/9409089]. 

\bibitem{SS}
S.~H.~Shenker and D.~Stanford, 
``Black holes and the butterfly effect,'' 
JHEP {\bf 1403}, 067 (2014)
[arXiv:1306.0622[hep-th]]. 

\bibitem{RS}
D.~A.~Roberts, D.~Stanford,
``Two-dimensional conformal field theory and the butterfly effect,''
Phys.Rev.Lett. 115 (2015) 13, 131603 
[arXiv:1412.5213[hep-th]]. 

\bibitem{MSS}
J.M.~Maldacena, S.~H.~Shenker and D.~Stanford,
``A bound on chaos,''
arXiv:1503.01409 [hep-th]. 

\bibitem{CSSTW}
P.~Caputa, J.~Simon, A.~Stikonas, T.~Takayanagi and K.~Watanabe,
``Scrambling time from local perturbations of the eternal BTZ black hole,''
JHEP 1508 (2015) 011
[arXiv:1503.08161 [hep-th]]. 

\bibitem{Miyaji}
M.~Miyaji,
``Butterflies from Information Metric,''
JHEP {\bf1609} (2016) 002,
[arXiv:1607.01467 [hep-th]]. 

\bibitem{LO}
A.~Larkin and Y.~Ovchinnikov, 
``Quasiclassical method in the theory of superconductivity,'' 
JETP 28,6 (1969) 1200-1205

\bibitem{Kitaev}
A.~Kitaev,
``A simple model of quantum holography,''
Talks at KITP, April 7,2015 and May 27, 2015.

\bibitem{Polchinski}
J.~Polchinski,
``Chaos in the black hole S-matrix,''
arXiv:1505.08108 [hep-th]. 

\bibitem{TV}
G.~Turiaci and H.~Verlinde,
``On CFT and Quantum Chaos,''
arXiv:1603.03020 [hep-th]. 

\bibitem{Perlmutter}
E.~Perlmutter,
``Bounding the Space of Holographic CFTs with Chaos,''
arXiv:1602.08272 [hep-th]. 

\bibitem{CNO}
P.~Caputa, T.~Numasawa and A.~V.~Osorio,
``Scrambling without chaos in RCFT,''
arXiv:1602.06542 [hep-th]. 

\bibitem{GQ}
Y.~Gu and X.~L.~Qi,
``Fractional Statistics and the Butterfly Effect,''
JHEP {\bf1608} (2016) 129  
[arXiv:1602.06543 [hep-th]]. 


\bibitem{KMS}
J.~L.~Karczmarek, J.~M.~Maldacena and A.~Strominger,
"Black hole non-formation in the matrix model,"
JHEP {\bf0601} (2006) 039 
[arXiv:hep-th/0411174 ]. 

\bibitem{MaSt}
J.~M.~Maldacena and D.~Stanford,
"Comments on the Sachdev-Kitaev-Ye model,"
arXiv:1604.07818 [hep-th]. 




\bibitem{CC1}
P.~Calabrese and J.~Cardy,
``Evolution of Entanglement Entropy in One-Dimensional Systems,''
J. Stat. Mech. {\bf0504} (2005) P04010 
[arXiv:cond-mat/0503393 ]. 

\bibitem{ABGH1}
C.~T.~Asplund, A.~Bernamonti, F.~Galli, and T.~Hartman,
``Entanglement Scrambling in 2d conformal Field Theory,''
JHEP {\bf1509} (2015) 110 
[arXiv:1506.03772 [hep-th]]. 




\bibitem{NNT}
M.~Nozaki, T.~Numasawa, T.~Takayanagi,
``Quantum Entanglement of Local Operators in Conformal Field Theories,''
Phys. Rev. Lett. 112, 111602 (2014) 
[arXiv:1401.0539 [hep-th]]. 


\bibitem{HNTW}
S.~He, T.~Numasawa, T.~Takayanagi, K.~Watanabe,
``Quantum Dimension as Entanglement Entropy in 2D CFTs,''
Phys. Rev. D 90, 041701 (2014) 
[arXiv:1403.0702 [hep-th]]. 

\bibitem{Nozaki}
M.~Nozaki,
``Notes on Quantum Entanglement of Local Operators,''
JHEP {\bf 1410} (2014) 147 
[arXiv:1405.5875 [hep-th]]. 

\bibitem{GH}  
W.~Z.~Guo and S.~He,
``R\'enyi entropy of locally excited states with thermal and boundary effect in 2D CFTs,''
JHEP {\bf 1504}, 099 (2015)
[arXiv:1501.00757 [hep-th]].

\bibitem{CNT}
P.~Caputa, M.~Nozaki and T.~Takayanagi,
``Entanglement of Local Operators in large N CFTs,''
PTEP 2014 (2014) 093B06 
[arXiv:1405.5946 [hep-th]]. 

\bibitem{ABGH2}
C.~T.~Asplund, A.~Bernamonti, F.~Galli, and T.~Hartman,ｔ
``Holographic Entanglement Entropy from 2d CFT: Heavy States and Local Quenches,''
JHEP {\bf 1502} (2015) 171 
[arXiv:1410.1392 [hep-th]]. 

\bibitem{CSST}
P.~Caputa, J.~Simon, A.~Stikonas and T.~Takayanagi,
``Quantum Entanglement of Localized Excited States at Finite Temperature,''
JHEP {\bf1501} (2015) 102 
[arXiv:1410.2287 [hep-th]]. 


\bibitem{CLM}
H.~Casini, H.~Liu and M.~Mezei,
``Spread of entanglement and causality,''
JHEP {\bf 1607} (2016) 077
[arXiv:1509.05044 [hep-th]]. 

\bibitem{CO}
P.~Caputa and A.~Veliz-Osorio,
``Entanglement constant for conformal families,''
Phys. Rev. D 92, 065010 (2015) 
[arXiv:1507.00582 [hep-th]]. 




\bibitem{CGHW}
B.~Chen,W.~Z.~Guo, S.~He and J.~Wu,
``Entanglement Entropy for Descendent Local operators in 2D CFTs,''
[arXiv:1507.01157 [hep-th]]. 



\bibitem{HJK}
T.~Hartman, S.~Jain and S.~Kundu,
``Causality Constraints in Conformal Field Theory,''
[arXiv:1509.00014 [hep-th]]. 

\bibitem{RT}
S.~Ryu and T.~Takayanagi,
``Aspects of Holographic Entanglement Entropy,''
JHEP 0608 (2006) 045 
[arXiv:hep-th/0605073 ]. 

\bibitem{Yellowbook}
 P.~Di Francesco, P.~Mathieu and D.S\'{e}n\'{e}chal, 
 {\it Conformal Field Theory}, Springer-Verlag, 1997.

\bibitem{CR}
P.~Caputa and M.~Rams,
``Quantum dimensions from local operator excitations in the Ising model,'' 
arXiv:1609.02428 [cond-mat.str-el]. 

\bibitem{MS}
G.~Moore and N.~Seiberg,
``Classical and Quantum conformal field theory,''
Commun.Math.Phys. {\bf123 } (1989) 177

\bibitem{GG}
M.~R.~Gaberdiel and R.~Gopakumar,
``Minimal Model Holography,''
J.Phys. A{\bf46} (2013) 214002
[arXiv:1207.6697 [hep-th]]. 

\bibitem{AHRS}
T.~Anous, T.~Hartman, A.~Rovai and J.~Sonner,
``Black Hole Collapse in the $1/c$ Expansion,''
JHEP {\bf1607} (2016) 123 
[arXiv:1603.04856 [hep-th]]. 


\bibitem{NNTlocal}
M.~Nozaki, T.~Numasawa and T.~Takayanagi,
``Holographic Local Quenches and Entanglement Density,''
 JHEP {\bf 1305} (2013) 080 
[arXiv:1302.5703 [hep-th]].


\bibitem{FGMP}
J.~J.~Friess, S.~S.~Gubser, G.~Michalogiorgakis and S.~S.~Pufu,
``Expanding plasmas and quasinormal modes of anti-de Sitter black holes,''
JHEP {\bf0704} (2007) 080 
[arXiv:hep-th/0611005]. 

\bibitem{FHRR}
P.~Figueras, V.~E.~Hubeny, M.~Rangamani and S.~F.~Ross,
``Dynamical black holes and expanding plasmas,''
JHEP {\bf 0904} (2009) 137
[arXiv:0902.4696 [hep-th]]. 



\end{thebibliography}
\end{document}